\documentclass[twocolumn]{aastex631}

\usepackage{amsmath}

\begin{document}

\title{Flipping of the tidal tails of the Ophiuchus stream due to the decelerating Galactic bar}

\correspondingauthor{Yong~Yang}
\email{yong.yang@sydney.edu.au}

\author[0000-0001-7609-1947]{Yong~Yang}
\affiliation{Sydney Institute for Astronomy, School of Physics, A28, The University of Sydney, NSW 2006, Australia}

\author[0000-0003-3081-9319]{Geraint~F.~Lewis}
\affiliation{Sydney Institute for Astronomy, School of Physics, A28, The University of Sydney, NSW 2006, Australia}

\author[0000-0002-8448-5505]{Denis~Erkal}
\affiliation{School of Mathematics and Physics, University of Surrey, Guildford GU2 7XH, UK}

\author[0000-0002-9110-6163]{Ting~S.~Li}
\affiliation{Department of Astronomy \& Astrophysics, University of Toronto, 50 St. George Street, Toronto ON, M5S 3H4, Canada}
\affiliation{Dunlap Institute for Astronomy \& Astrophysics, University of Toronto, 50 St George Street, Toronto, ON M5S 3H4, Canada}
\affiliation{Data Sciences Institute, University of Toronto, 17th Floor, Ontario Power Building, 700 University Ave, Toronto, ON M5G 1Z5, Canada}

\author[0009-0005-5355-5899]{Andrew~P.~Li}
\affiliation{Department of Astronomy \& Astrophysics, University of Toronto, 50 St. George Street, Toronto ON, M5S 3H4, Canada}

\author[0000-0002-3430-4163]{Sarah~L.~Martell}
\affiliation{School of Physics, University of New South Wales, Sydney, NSW 2052, Australia}
\affiliation{Centre of Excellence for All-Sky Astrophysics in Three Dimensions (ASTRO 3D), Australia}

\author[0000-0001-8536-0547]{Lara~R.~Cullinane}
\affiliation{Leibniz-Institut f{\"u}r Astrophysik Potsdam (AIP), An der Sternwarte 16, D-14482 Potsdam, Germany}

\author[0000-0002-9269-8287]{Guilherme~Limberg}
\affiliation{Kavli Institute for Cosmological Physics, University of Chicago, Chicago, IL 60637, USA}

\author[0000-0003-1124-8477]{Daniel~B.~Zucker}
\affiliation{School of Mathematical and Physical Sciences, Macquarie University, Sydney, NSW 2109, Australia}
\affiliation{Macquarie University Research center for Astrophysics and Space Technologies, Sydney, NSW 2109, Australia}
\affiliation{Centre of Excellence for All-Sky Astrophysics in Three Dimensions (ASTRO 3D), Australia}

\author[0000-0001-7516-4016]{Joss~Bland-Hawthorn}
\affiliation{Sydney Institute for Astronomy, School of Physics, A28, The University of Sydney, NSW 2006, Australia}
\affiliation{Centre of Excellence for All-Sky Astrophysics in Three Dimensions (ASTRO 3D), Australia}

\author[0000-0002-6021-8760]{Andrew~B.~Pace}
\affiliation{Department of Astronomy, University of Virginia, 530 McCormick Road, Charlottesville, VA 22904, USA}

\author[0000-0001-7019-649X]{Gary~S.~Da~Costa}
\affiliation{Research School of Astronomy and Astrophysics, Australian National University, Canberra, ACT 2611, Australia}
\affiliation{Centre of Excellence for All-Sky Astrophysics in Three Dimensions (ASTRO 3D), Australia}

\author[0000-0002-4863-8842]{Alexander~P.~Ji}
\affiliation{Department of Astronomy \& Astrophysics, University of Chicago, 5640 S Ellis Avenue, Chicago, IL 60637, USA}
\affiliation{Kavli Institute for Cosmological Physics, University of Chicago, Chicago, IL 60637, USA}

\author[0000-0003-2644-135X]{Sergey~E.~Koposov}
\affiliation{Institute for Astronomy, University of Edinburgh, Royal Observatory, Blackford Hill, Edinburgh EH9 3HJ, UK}
\affiliation{Institute of Astronomy, University of Cambridge, Madingley Road, Cambridge CB3 0HA, UK}

\author[0000-0003-0120-0808]{Kyler~Kuehn}
\affiliation{Lowell Observatory, 1400 W Mars Hill Rd, Flagstaff,  AZ 86001, USA}

\author[0000-0003-2497-091X]{Nora~Shipp}
\affiliation{Department of Astronomy, University of Washington, Seattle, WA 98195, USA}

\author[0009-0003-7075-3235]{Miles~Pearson}
\affiliation{School of Physics, University of New South Wales, Sydney, NSW 2052, Australia}

\author[0000-0003-0918-7185]{Sam~A.~Usman}
\affiliation{Department of Astronomy \& Astrophysics, University of Chicago, 5640 S Ellis Avenue, Chicago, IL 60637, USA}
\affiliation{Kavli Institute for Cosmological Physics, University of Chicago, Chicago, IL 60637, USA}

\author{$S^5$~Collaboration}



\begin{abstract}
The Ophiuchus stellar stream presents a puzzle due to its complicated morphology, with a substructure perpendicular to the main track (spur), a broadened tail (fanning), and a shorter than expected angular extent given its old stellar population and short orbital period. The location of the stream approaches the Galactic center, implying a possible connection between its orbit and its unusual morphology. Here we demonstrate that the morphology of Ophiuchus can be attributed to its interaction with the decelerating Galactic bar, which leads to the flipping or transposition of its tidal tails. The short length of the stream is the result of stars stripped in the ancient past still remaining concentrated, and the spur, as well as the fanning, are composed of either leading or trailing tails built up of stars released at different time intervals. Our new spectroscopic data, obtained as part of the Southern Stellar Stream Spectroscopic Survey $(S^5)$, and modeling of Ophiuchus indicate that, in the presence of the bar, an initial leading tail can be redistributed to the trailing side and vice versa, and the morphology of a stream can be reshaped. This result confirms that the Galactic bar plays a vital role in reconstructing the orbital behavior of streams passing close to the central region of the Milky Way.
\end{abstract}

\keywords{Stellar streams(2166) --- Globular star clusters(656) --- Galactic bar(2365) --- Milky Way Galaxy(1054) --- Stellar dynamics(1596) --- Milky Way dynamics(1051)}


\section{Introduction} \label{sec:intro}

Galactic stellar streams are arc-like structures formed when globular clusters or dwarf galaxies are tidally disrupted by the Milky Way gravitational field \citep{2006ApJ...643L..17G,2016MNRAS.463.1759B,2018ApJ...862..114S,2020Natur.583..768W,2021ApJ...914..123I,2022ApJ...935L..38Y,2023ApJ...945L...5Y}. They are sensitive to gradients in the Galactic gravitational potential \citep{2002ApJ...570..656J,2002MNRAS.332..915I}, making them valuable tools for understanding the structure and evolution of the Milky Way, especially the morphology of the extensive dark matter distribution \citep{2024ApJ...967...89I,2019MNRAS.487.2685E,2023MNRAS.521.4936K,2022ApJ...926..107M,2025NewAR.10001713B}. 

The effects from the Galactic bar on stellar streams have already been identified and discussed \citep{2017MNRAS.470...60E,2017NatAs...1..633P}. The Ophiuchus stellar stream \citep{2014MNRAS.443L..84B}, a short narrow stream generated by a globular cluster \citep{2022ApJ...928...30L} discovered near the Galactic center, stands out as another intriguing case. Its location makes it an ideal probe for exploring the intricate interplay between the gravitational influence of the stream dynamics and the central bar/bulge potential \citep{2016ApJ...824..104P,2016MNRAS.460..497H}. The stream provides crucial information about the central bar, e.g. the strength of the bar's torque today, in a region that is difficult to study because of the projected high stellar densities and dust obscuration.

Models that adopt an axisymmetric Galactic potential suggest that the Ophiuchus progenitor began disrupting relatively recently, starting roughly 300 Myr ago \citep{2015ApJ...809...59S,2020MNRAS.492.4164L}, to match its short deprojected length of 1.5 kpc \citep{2015ApJ...809...59S}. Although the stream is now known to be twice as long \citep{2020AJ....159..287C}, the disruption timescale increases to only $\sim$ 600 Myr in this scenario. However, since the stream has a stellar population of around 12 Gyr old \citep{2015ApJ...809...59S} with an orbital period of only $\sim$ 200 Myr \citep{2016ApJ...824..104P,2020AJ....159..287C,2022ApJ...928...30L},
it is more likely that the disruption began long ago rather than merely a few dynamical times in the past. Furthermore, the stream possesses quite low energy and is deeply buried in the Galactic gravitational potential well \citep[e.g., Figure 3 in][]{2022ApJ...928...30L}. It seems unlikely that the Ophiuchus system merged with the Milky Way at a recent time. Just when the progenitor system moved in close enough within the gravitational field to undergo disruption \citep[e.g.,][]{2000MNRAS.314..468J} is very uncertain. Although this puzzle motivates further study, we do not discuss the origin and survival of the stream in the present work. Our focus is on how the stream usefully probes the center of the Galaxy.

By incorporating the Galactic bar, there are two hypotheses to explain the short length of Ophiuchus: chaotic fanning \citep{2016ApJ...816L...4S,2016ApJ...824..104P} and bar shepherding \citep{2016MNRAS.460..497H}. The former is a phenomenon where the stream shows abrupt broadening or spreading in both position and velocity space, i.e., an increase in the width and velocity dispersion. The time-dependent bar induces chaotic orbits, dispersing the stream stars and making it appear shorter. The latter is a mechanism where differential torques along the stream from the bar's rotating potential can reshape the stream's energy distribution and slow down its growth rate, allowing it to remain short for a longer period. In addition to the short length and fanning of the stream, the stream also displays a ``spur'' feature \citep{2020AJ....159..287C}, that is, there are stars extending outward from the main linear body of the stream. One commonly adopted interpretation for a spur in a stream is an encounter with a dark subhalo that can gravitationally perturb member stars, causing some to deviate from the main stream track and form a separate branch \citep{2019ApJ...880...38B,2022ApJ...941..129D}; this further implies a dynamically complex history of the Ophiuchus stream. 

Here, we describe a scenario that explains all of the unusual characteristics of the Ophiuchus stream. We use a Bayesian Mixture Model \citep[][A. Li et al. in prep.]{2019ApJ...885....3S,2025A&A...693A..69A} to select the stream member stars using data from the Southern Stellar Stream Spectroscopic Survey \citep[$S^5$;][]{2019MNRAS.490.3508L}. By adopting a realistic Milky Way potential \citep{2024A&A...692A.216H} including a decelerating bar \citep{2022MNRAS.514L...1S}, we can reproduce the short length, fanning, and spur features simultaneously in agreement with observation. The simulation also reveals an intriguing evolutionary history of the stream.

We organize the paper as follows:
In Section~\ref{sec:data} we introduce the $S^5$ data. In Section~\ref{sec:methods} we describe the Bayesian Mixture Model used to identify the Ophiuchus members, and detail the simulation setups in modeling the stream. Section~\ref{sec:results} presents our main findings of Ophiuchus based on simulations. In Section~\ref{sec:summary} we summarize this work. 

\section{Data} \label{sec:data}

The Southern Stellar Stream Spectroscopic Survey \citep[$S^5$;][]{2019MNRAS.490.3508L} is aimed at obtaining deep insights into known streams in the southern hemisphere. Observations were initiated in 2018 with the Two-degree Field (2dF) fiber positioner \citep{2002MNRAS.333..279L} coupled with the dual-arm AAOmega spectrograph on the 3.9-m Anglo-Australian Telescope (AAT). 

For the Ophiuchus stream, 8 AAT fields were observed over 2020-2021. The data catalog used in this analysis is from an internal data release, iDR3.7, which will be made public as the second public data release for $S^5$ in 2025 (T. S. Li et al. in prep.). The catalog provides spectroscopic measurements for more than two thousand candidate members in the Ophiuchus fields, including radial velocity and metallicity measurements as determined by the \texttt{rvspecfit} pipeline \citep{2019ascl.soft07013K}. Proper motions in Right Ascension (R.A.) and declination (Dec.), $\mu^*_\alpha$ and $\mu_\delta$, along with their covariances, $\rho$, are cross-matched from Gaia DR3 \citep{2023A&A...674A...1G}. 

We briefly describe here the spectroscopic target selection function in source selection of $S^5$ Ophiuchus observations, which is used in the model dispersion measurement in Section~\ref{subsec:comparison}. This includes selections on sky coverage and proper motions. There are eight footprints in the stream field centered at (R.A., Dec.) = (240.00, $-7.25$)$^\circ$, (240.40, $-5.74$)$^\circ$, (241.80, $-6.87$)$^\circ$,  (242.10, $-5.54$)$^\circ$, (243.60, $-6.80$)$^\circ$, (245.40, $-7.03$)$^\circ$, (247.20, $-7.57$)$^\circ$, and (249.00, $-8.41$)$^\circ$, each with a radius of 1.05$^\circ$ (field-of-view of the 2dF fiber positioner). Candidate sources are further constrained using preliminary proper motion tracks. Specifically, we derive a $\mu^*_\alpha$ track and a constant $\mu_\delta$ track with member stars in \citet{2020AJ....159..287C}, and compute residuals between proper motions and tracks as follows:
\begin{equation}
\begin{aligned}
& \mu^*_{\alpha,{\rm res}} = \mu^*_\alpha - \\ 
& (4.7801\times10^{-2}{\rm R.A.}^2 -24.2144{\rm R.A.} + 3.0552\times10^3)
\end{aligned}
\end{equation}
and $\mu_{\delta,{\rm res}}$ = $\mu_\delta$ + 4.5 (placing them around zero). A criterion of $|\mu^*_{\alpha,{\rm res}}|$ and $|\mu_{\delta,{\rm res}}|$ both $<$ 1.5 mas yr$^{-1}$ is then applied.

\section{Methods} \label{sec:methods}

\subsection{Bayesian Mixture Model} \label{subsec:bayesian}

To simplify the stream member selection, we define stream-aligned coordinate system ($\phi_1$, $\phi_2$) (parallel and perpendicular to the stream) by specifying two endpoints (R.A., Dec.) = (246.6$^\circ$, $-7.7^\circ$) and (242.3$^\circ$, $-6.9^\circ$) and rotating from the ICRS to the frame of a great circle that passes through the endpoints. This is achieved with the \texttt{astropy} \citep{2013A&A...558A..33A} and \texttt{gala} \citep{2017JOSS....2..388P} python packages. We also convert heliocentric radial velocities ($V_r$) to the Galactic standard of rest ($V_{\rm gsr}$).

\begin{figure*}[htb!]
\centering
\includegraphics[width=\textwidth]{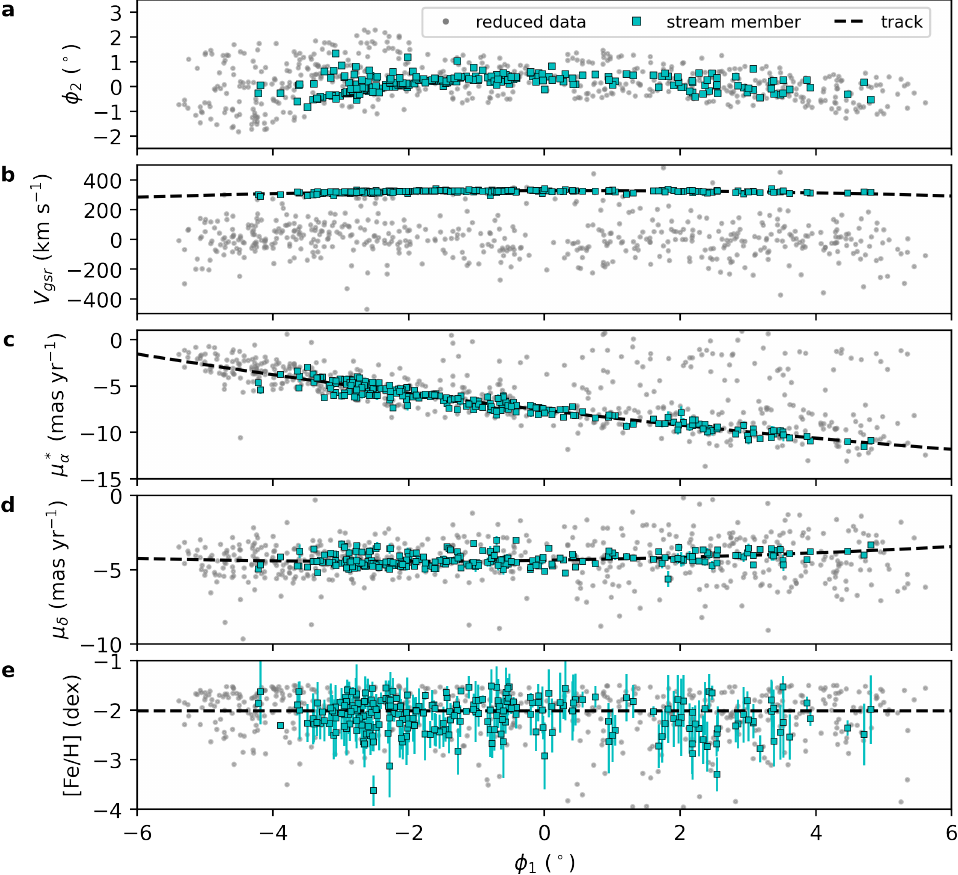}
\caption{Selection of the Ophiuchus member stars. Gray dots show 747 stars passing the criteria applied to the initial $S^5$ data. Cyan squares represent ultimately adopted Ophiuchus members with probability $>$ 0.9. Dashed lines indicate the best fit tracks to the stream using the Bayesian Mixture Model. From panel \textbf{a} to \textbf{e} are $\phi_2$, $V_{\rm gsr}$, $\mu^*_\alpha$, $\mu_\delta$, and [Fe/H] as a function of $\phi_1$.}
\label{fig:member}
\end{figure*}

The following criteria are first applied to the catalog to remove stars far from the Ophiuchus stream and/or to ensure the quality of the velocity and metallicity measurements:
\begin{itemize}
    \item metallicity [Fe/H] $<$ $-$1.5 dex,
    \item $-$500 km s$^{-1}$ $<$ $V_{\rm gsr}$ $<$ 500 km s$^{-1}$,
    \item $-$15 mas yr$^{-1}$ $<$ $\mu^*_\alpha$ $<$ 1 mas yr$^{-1}$,
    \item $-$10 mas yr$^{-1}$ $<$ $\mu_\delta$ $<$ 0 mas yr$^{-1}$,
    \item uncertainty of $V_r$ $<$ 20 km s$^{-1}$,
    \item uncertainty of [Fe/H] $<$ 1 dex, and
    \item flag \texttt{good\_star} = 1,
\end{itemize}
where \texttt{good\_star} is a derived parameter for $S^5$ data used to remove poor spectroscopic fits or non-stellar objects \citep{2019MNRAS.490.3508L}. As a result, the initial catalog of 2,666 stars is reduced to 747 stars (gray dots in Figure~\ref{fig:member}).

A Bayesian Mixture Model \citep[][A. Li et al. in prep.]{2019ApJ...885....3S,2025A&A...693A..69A} is used to determine the probability of a star being a stream member. In brief, the mixture model consists of two stellar components, a stream and a background. A parameter $f_s$ (the subscript ``s'' denotes ``stream'') is defined to account for the fraction of the stream stars from all candidates. At any given $\phi_1$ of input candidates, the mixture Gaussians are constructed. The stream component is represented by a one-dimensional Gaussian in $V_{\rm gsr}$ and [Fe/H], and a two-dimensional Gaussian in proper motions. For these Gaussians, the mean in $V_{\rm gsr}$ and proper motion are allowed to vary as a second-order polynomial of $\phi_1/10^\circ$, i.e.,
\begin{equation}
m_{k,s} = A_{k,s} + B_{k,s} \phi_1/10^\circ + C_{k,s} (\phi_1/10^\circ)^2
\end{equation}
where $k$ = $V_{\rm gsr}$, $\mu^*_\alpha$, and $\mu_\delta$, to account for the kinematic changes along the stream. The mean of [Fe/H] is kept constant $m_{{\rm [Fe/H]},s}$. For Gaussian variances, we use the quadratic sum of intrinsic dispersions of the stream $\sigma_{k,{\rm intr}}$ (to be determined by the model) and the measured data uncertainties $\sigma_{k,{\rm meas}}$, that is
\begin{equation}
\sigma^2_{k,s} = \sigma^2_{k,{\rm intr}} + \sigma^2_{k,{\rm meas}} .
\end{equation}
For proper motions, we further take into account the correlation $\rho$ between $\sigma_{\mu^*_\alpha,{\rm meas}}$ and $\sigma_{\mu_\delta,{\rm meas}}$:
\begin{equation}
\Sigma_s = 
\begin{bmatrix}
\sigma^2_{\mu^*_\alpha,{\rm intr}} + \sigma^2_{\mu^*_\alpha,{\rm meas}} &\: & 
\rho \sigma_{\mu^*_\alpha,{\rm meas}} \sigma_{\mu_\delta,{\rm meas}} \\ 
\rho \sigma_{\mu^*_\alpha,{\rm meas}} \sigma_{\mu_\delta,{\rm meas}} &\: & 
\sigma^2_{\mu_\delta,{\rm intr}} + \sigma^2_{\mu_\delta,{\rm meas}} 
\end{bmatrix} .
\end{equation}
Given a star $i$ with $V_{\rm gsr}$, [Fe/H], $\mu^*_\alpha$, and $\mu_\delta$, the likelihood of the stream component in each aspect can be given by
\begin{equation}
\mathcal{N}^i_{V_{\rm gsr},s} = 
\frac{1}{\sigma_{V_{\rm gsr},s}\sqrt{2\pi}} {\rm exp}
\left[
-\frac{(V_{\rm gsr} - m_{V_{\rm gsr},s})^2}{2\sigma^2_{V_{\rm gsr},s}}
\right]
\end{equation}

\begin{equation}
\mathcal{N}^i_{{\rm [Fe/H]},s} = 
\frac{1}{\sigma_{{\rm [Fe/H]},s}\sqrt{2\pi}} {\rm exp}
\left[
-\frac{({\rm [Fe/H]} - m_{{\rm [Fe/H]},s})^2}{2\sigma^2_{{\rm [Fe/H]},s}}
\right]
\end{equation}

\begin{equation}
\mathcal{N}^i_{\mu,s} = 
\frac{1}{\sqrt{|\Sigma_s|}2\pi} {\rm exp} 
\left( - \frac{1}{2}
\begin{bmatrix}
\mu^*_\alpha - m_{\mu^*_\alpha,s} \\
\mu_\delta - m_{\mu_\delta,s}
\end{bmatrix}^T\Sigma^{-1}_s
\begin{bmatrix}
\mu^*_\alpha - m_{\mu^*_\alpha,s} \\
\mu_\delta - m_{\mu_\delta,s}
\end{bmatrix}
\right)
\end{equation}
As for the background component, the setting is nearly the same as above, except that Gaussian means are chosen to be constant having no $\phi_1$ dependence -- i.e., $m_{k,b}$ = $A_{k,b}$ (subscript ``b'' denotes ``background'') -- as the short stream length means this is a reasonable assumption. Finally, the possibility of observing star $i$ can be written as 
\begin{equation}
\mathcal{L}^i = f_s \mathcal{N}^i_{V_{\rm gsr},s} \mathcal{N}^i_{{\rm [Fe/H]},s} 
\mathcal{N}^i_{\mu,s} + (1-f_s) 
\mathcal{N}^i_{V_{\rm gsr},b} \mathcal{N}^i_{{\rm [Fe/H]},b} \mathcal{N}^i_{\mu,b}
\end{equation}
and the total likelihood for all stars is
\begin{equation}
\mathcal{L} = \prod_i \mathcal{L}^i 
\end{equation}
In total, there are 23 parameters to be fitted, including the stream fraction (1), coefficients of polynomials describing kinematic changes along the stream (9), the stream metallicity (1), relevant intrinsic dispersions (4), as well as means (4) and dispersions (4) for the background component. We assume a log-flat prior for the intrinsic dispersions, and a flat prior for the other parameters. Inputs to the posterior function are an initial guess of the 23 parameters, $\phi_1$, the line-of-sight velocities, metallicities, proper motions, and all associated measurement uncertainties of data. We sample the posterior with a Markov Chain Monte Carlo (MCMC) method built in the \texttt{emcee} python package \citep{2013PASP..125..306F}. The medians along each dimension serve as the best-fit values.

After obtaining these parameters, they are substituted within the above Gaussians, and the membership probability of each star is calculated by
\begin{equation}
\mathcal{P}^i = \frac{
f_s \mathcal{N}^i_{V_{\rm gsr},s} \mathcal{N}^i_{{\rm [Fe/H]},s} 
\mathcal{N}^i_{\mu,s}  
}{
f_s \mathcal{N}^i_{V_{\rm gsr},s} \mathcal{N}^i_{{\rm [Fe/H]},s} 
\mathcal{N}^i_{\mu,s} + (1-f_s) 
\mathcal{N}^i_{V_{\rm gsr},b} \mathcal{N}^i_{{\rm [Fe/H]},b} \mathcal{N}^i_{\mu,b}
} .
\end{equation}
We consider stars with $\mathcal{P}^i$ $>$ 0.9 to be Ophiuchus members, of which there are 237, and use these  in the analysis of this work. Figure~\ref{fig:member} shows these 237 member stars (cyan squares) as well as the best-fit tracks of the stream component using the above Bayesian Mixture Model. One thing worth noting is that the mixture model identifies stream stars depending only on metallicity and kinematics as a function of $\phi_1$, not $\phi_2$, which allows us to unveil the complicated spatial morphology of the stream, such as the spur. The identified member stars are tabulated in Table~\ref{tab:star237}.

\begin{table*}[htb!]
\scriptsize
\caption{The Ophiuchus stream member stars (probability $>$ 0.9) identified with the Bayesian Mixture Model from $S^5$ data}
\label{tab:star237}
\begin{center}
\begin{tabular}{c c c c c c c c c c c}
\hline
\hline
Gaia ID & R.A. & Dec. & $\mu^*_\alpha$ & $\mu_\delta$ & $G_0$ & $(G_{\rm BP}-G_{\rm RP})_0$ & $V_r$ & $\sigma_{V_r,{\rm meas}}$ & [Fe/H] & e[Fe/H] \\
 &  $^\circ$ & $^\circ$ & mas yr$^{-1}$ & mas yr$^{-1}$ & mag & mag & km s$^{-1}$ & km s$^{-1}$  & dex  & dex \\
\hline
4349192941740794624  &  242.7455  &   -6.7556  &   -6.024  &   -4.952  &  18.74  &   0.57  &  289.56 &    5.61 &   -2.488  &  0.531 \\
4349552211464342784  &  241.3988  &   -7.1038  &   -4.255  &   -4.560  &  16.79  &   0.89  &  281.22 &    1.13 &   -1.839  &  0.073 \\
4350170751174810112  &  242.6808  &   -5.7643  &   -7.122  &   -3.812  &  19.07  &   0.59  &  270.33 &   12.30 &   -2.364  &  0.543 \\
4350716693057504384  &  247.8398  &   -8.2642  &   -9.815  &   -3.629  &  18.84  &   0.66  &  275.72 &   10.82 &   -2.448  &  0.554 \\
4349058457723585792  &  243.7409  &   -6.8564  &   -7.551  &   -4.410  &  18.63  &   0.71  &  294.83 &    4.65 &   -2.678  &  0.296 \\
4349193972532692096  &  242.4450  &   -6.9499  &   -5.133  &   -4.524  &  18.32  &   0.59  &  287.83 &    4.84 &   -1.702  &  0.360 \\
4349232081779281792  &  243.2040  &   -6.8615  &   -6.481  &   -4.337  &  16.54  &   0.88  &  291.28 &    0.89 &   -1.893  &  0.060 \\
4349920822739660416  &  241.7780  &   -6.9709  &   -4.936  &   -4.666  &  18.59  &   0.55  &  280.78 &   11.84 &   -2.635  &  0.244 \\
4349941090682710656  &  241.8759  &   -6.8270  &   -5.358  &   -4.602  &  18.19  &   0.60  &  291.71 &    3.84 &   -2.327  &  0.267 \\
4351938490993912576  &  245.5212  &   -7.2830  &   -8.221  &   -4.041  &  18.60  &   0.67  &  277.73 &   15.38 &   -2.412  &  0.311 \\
\hline
\end{tabular}
\end{center}
\tablecomments{The columns from left to right correspond to the following: Gaia DR3 ID, celestial positions, proper motions, magnitude, color, $S^5$ heliocentric radial velocity and its error, metallicity and its error.\\
Table 1 is published in its entirety in the machine-readable format. A portion is shown here for guidance regarding its form and content.}
\end{table*}

\subsection{The barred Milky Way potential} \label{subsec:pot}
N-body simulations suggest that central galactic bars will typically slow down due to angular momentum transfer with the dark matter halo \citep{2000ApJ...543..704D,2015MNRAS.454.3166A,2019MNRAS.488.4552S,2021MNRAS.500.4710C}. Therefore, unlike previous studies of the Ophiuchus stream, where a bar rotating at constant speed is used \citep{2016ApJ...824..104P,2016MNRAS.460..497H}, here we further consider the deceleration of the bar's pattern speed at a constant rate. 

We adopt a Milky Way potential model \citep{2024A&A...692A.216H} incorporating a central bar \citep{2017MNRAS.465.1621P,2022MNRAS.514L...1S} as part of the \texttt{AGAMA} python package \citep{2019MNRAS.482.1525V}. The potential consists of multiple components: a supermassive black hole, a nuclear star cluster, a nuclear stellar disk, a dark matter halo (together represented by \texttt{Multipole} expansions), a Galactic bar, thin and thick stellar disks, and two gas disks (together represented by \texttt{CylSpline} expansions). Among all these components, the bar is uniquely non-axisymmetric.

\begin{figure*}[htb!]
\centering
\includegraphics[width=\textwidth]{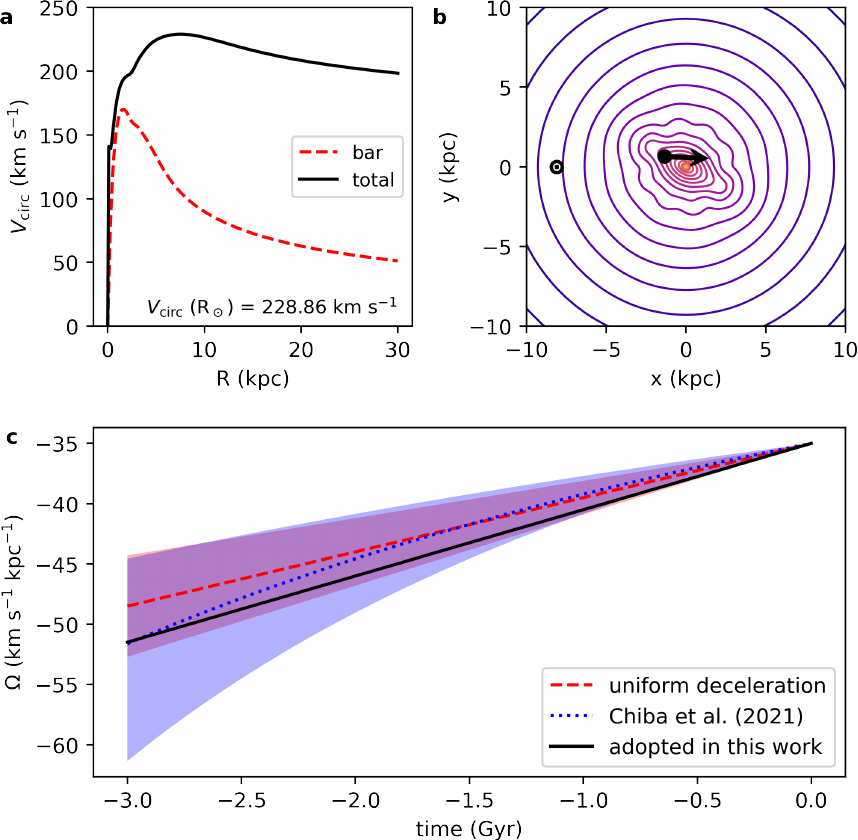}
\caption{The barred Milky Way potential. \textbf{a}, circular velocity as a function of radius for all Galactic components (black) and for the only bar (red). The value at the solar radius is 228.86 km s$^{-1}$. \textbf{b}, the surface density in the plane of the total potential at present time; $\odot$ stands for the Sun. The black point and arrow show the projected position and moving direction of Ophiuchus, which means the stream has a prograde polar orbit. \textbf{c}, comparing the pattern speed decelerating in different ways. The red shadow and dashed line denote the uniform deceleration $\Omega$ = $\Omega_0$ + $at$ with $a$ = 4.5 $\pm$ 1.4 km s$^{-1}$ kpc$^{-1}$ Gyr$^{-1}$ measured by \citet{2021MNRAS.500.4710C}. The blue shadow and dotted line correspond to $\Omega$ = $\Omega_0$ $/$ $(1-\eta \Omega_0 t)$ in \citet{2021MNRAS.500.4710C} with $\eta$ = 0.003 $\pm$ 0.001 \citep{2021MNRAS.500.4710C,2023MNRAS.525.3576C}. The black solid line is the adopted approach in our fiducial model. Here $\Omega_0$ = $-35$ km s$^{-1}$ kpc$^{-1}$.}
\label{fig:pot}
\end{figure*}

In the modeling, a right-hand Galactocentric Cartesian coordinate is used, where $x$ points from the Sun to the Galactic center, $y$ passes through the center in the direction of the Sun's orbital motion, and $z$ points towards the Galactic north pole. In such a frame, we describe rotation of the bar by assuming a uniformly accelerating formula 
\begin{equation}
\theta = \theta_0 + \Omega_0 t + \frac{1}{2} a t^2,
\label{eq:uniform}
\end{equation}
where the subscript ``0'' stands for present-day ($t$ = 0) values. $\theta_0$ and $\Omega_0$, the azimuthal angle of the bar's major axis with respect to the $x$ axis and the current pattern speed are chosen to be $-30^\circ$ and $-35$ km s$^{-1}$ kpc$^{-1}$ \citep{2016ARA&A..54..529B,2021MNRAS.500.4710C,2022ApJ...925...71L,2024MNRAS.533.3395Z}. The deceleration of the pattern speed, $a$, is adjusted to be 5.5 km s$^{-1}$ kpc$^{-1}$ Gyr$^{-1}$ within 1$\sigma$ of the measurement 4.5 $\pm$ 1.4 km s$^{-1}$ kpc$^{-1}$ Gyr$^{-1}$ from \citet{2021MNRAS.500.4710C}. The reason for adopting such a fiducial bar model is that the uniform deceleration is easy to construct, and, most importantly, it enables us to simulate the model stream in better agreement with the data morphologically. We compare the pattern speed changing in different ways over the past 3 Gyr in  Figure~\ref{fig:pot}, where it is seen that our fiducial model closely approximates the result of \citet{2021MNRAS.500.4710C} during this period. Figure~\ref{fig:pot} also displays the circular velocity as a function of radius for this barred potential along with the surface density contour if we look at the disk towards the $-z$ direction.

To transform between models and observations, we take the solar position as ($x$, $y$, $z$) = ($-$8.122, 0, 0.0208) kpc \citep{2018A&A...615L..15G,2019MNRAS.482.1417B} with velocity ($v_x$, $v_y$, $v_z$) = (12.9, 245.6, 7.78) km s$^{-1}$ \citep{2018RNAAS...2..210D}.

\subsection{Modeling the Ophiuchus stream} \label{subsec:model}
Observations reveal that the Ophiuchus stream has no extant core, implying that the globular cluster that generated the stream has dissolved completely. As \citet{2016ApJ...824..104P} pointed out, we also find that the stream's morphology can vary significantly between orbits even though these orbits are close to each other (i.e., Ophiuchus-like orbits are chaotic); this makes orbit fitting to the stream quite uncertain. $S^5$ combined with Gaia provides us with full five-dimensional measurements of all member stars, including R.A., Dec., $\mu^*_\alpha$, $\mu_\delta$, and $V_r$. Therefore, we start directly from observation by calculating data medians and combining them with the average distance from \citet{2022ApJ...928...30L} to represent the current orbit of the postulated progenitor. The final adopted orbit is (R.A., Dec., $d_\odot$, $\mu^*_\alpha$, $\mu_\delta$, $V_r$) = (243$^\circ$, $-6.895^\circ$, 7.9 kpc, $-6.3$ mas yr$^{-1}$, $-4.434$ mas yr$^{-1}$, 286 km s$^{-1}$).

The progenitor's potential is described as a Plummer sphere \citep{1911MNRAS..71..460P} with a mass $M_{\rm prog.}$ = 2 $\times$ 10$^4$ $M_\odot$ \citep{2015ApJ...809...59S} and a scale radius = 30 pc \citep{2016MNRAS.460..497H}. Its internal velocity dispersion $\sigma_v$ is assumed to be 0.8 km s$^{-1}$, smaller than our measurement in Figure~\ref{fig:model_vs_data}h, considering that the stream might have been dynamically heated by the bar. This value is chosen such that dispersions of the final model stream roughly match the observed dispersions. We apply a Lagrange-point-stripping method \citep{2014MNRAS.445.3788G,2019MNRAS.487.2685E} to generate the model stream. Particles are ejected at Lagrange points $r_{\rm prog.}$ $\pm$ $r_{\rm tidal}$ with random velocities $\mathbf{v}_{\rm prog.}$ $+$ $\mathcal{N}_3$(0, $\sigma_v$) every time step; $r_{\rm prog.}$ is the progenitor's distance to the Galactic center, and $r_{\rm tidal}$ is the tidal radius computed by
\begin{equation}
r_{\rm tidal} = \left( 
\frac{ G M_{\rm prog.} }{ \omega^2 - \frac{ \mathrm{d}^2\Phi }{ \mathrm{d}r^2_{\rm prog.} } } \right)^{ \frac{1}{3} },
\label{eq:rt}
\end{equation}
where
\begin{equation}
\omega = \frac{ |\mathbf{r}_{\rm prog.} \times \mathbf{v}_{\rm prog.}| }{ r^2_{\rm prog.} },
\end{equation}
and $G$ is the gravitational constant, $\Phi$ represents the Milky Way potential, and $\mathbf{r}_{\rm prog.}$ and $\mathbf{v}_{\rm prog.}$ are position and velocity vectors of the progenitor. $\mathcal{N}_3$(0, $\sigma_v$) is a three-dimensional Gaussian distribution with a dispersion of $\sigma_v$ in each dimension. We first integrate the progenitor in 1 Myr time steps back to 3 Gyr ago under the time-dependent potential (the bar rotates backwards following Equation~\ref{eq:uniform}). Here 3 Gyr is chosen as the integration time such that the model stream best matches the observations. The progenitor also happens to be at a pericenter at that time. From $-3$ Gyr to now, particles are released every 1 Myr and evolve under the combined potential of the progenitor and the Milky Way. We also let $M_{\rm prog.}$ linearly decrease to zero at the present day to mimic the mass loss process, and this is also applied when calculating $r_{\rm tidal}$ in Equation~\ref{eq:rt}.

\section{Results} \label{sec:results}

\subsection{Comparison between the data and model} \label{subsec:comparison}
\begin{figure}[htb!]
\centering
\includegraphics[width=\columnwidth]{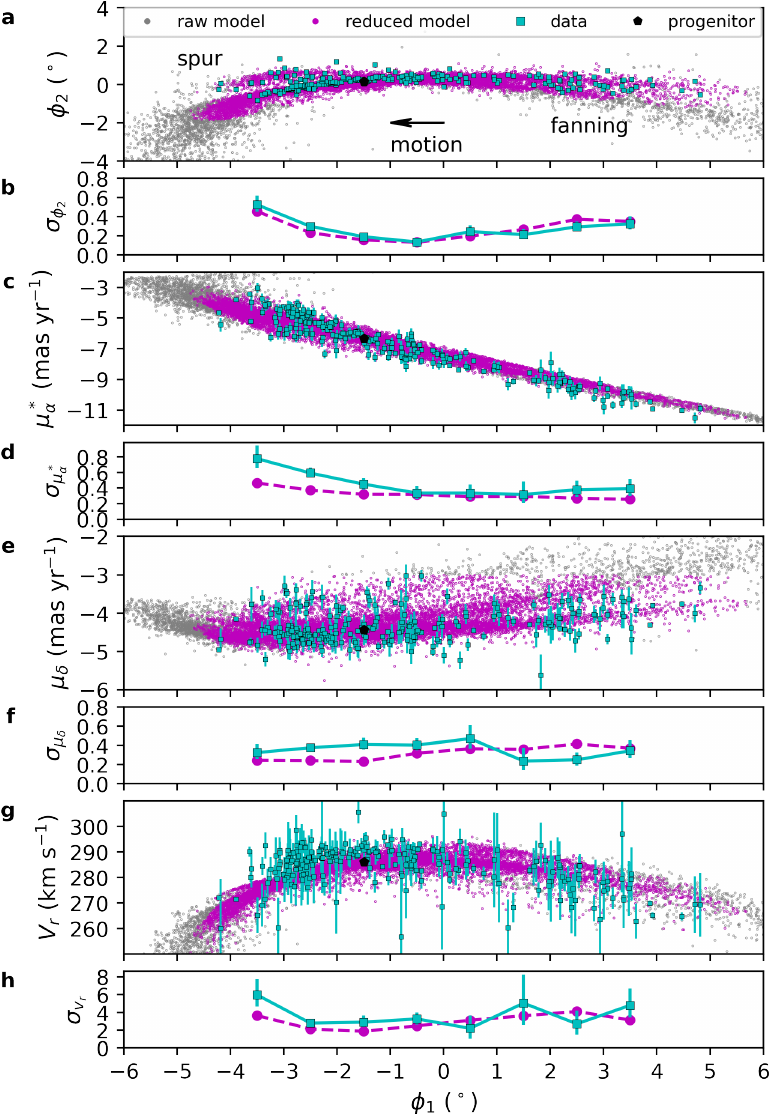}
\caption{Comparison in phase space between data and model. The postulated progenitor and the Ophiuchus stars are plotted with the black pentagon and cyan squares, respectively. Grey and magenta dots both represent our model particles, with magenta points the subset remaining after applying the spectroscopic selection function (see Section~\ref{sec:data}) to mimic the real observations. \textbf{a}, celestial position in stream-aligned coordinates ($\phi_1$, $\phi_2$). The black arrow roughly shows the direction of motion of the stream. \textbf{b}, dispersion in $\phi_2$ (width) as a function of $\phi_1$, estimated from magenta particles. The subsequent panel pairs similarly show changes along $\phi_1$ of proper motion in Right Ascension $\mu^*_\alpha$ (\textbf{c}), proper motion in declination $\mu_\delta$ (\textbf{e}), and radial velocity $V_r$ (\textbf{g}), as well as their associated dispersions (\textbf{d}, \textbf{f}, and \textbf{h}).}
\label{fig:model_vs_data}
\end{figure}

Figure~\ref{fig:model_vs_data} presents a comparison between member stars and model particles, showing position, proper motion, heliocentric radial velocity, and respective dispersions along the length of the Ophiuchus stream. Here, dispersions of the model are based on those particles (magenta) after applying the same spectroscopic target selection function (see details in Section~\ref{sec:data}), in order to imitate the real $S^5$ observations of the Ophiuchus stream. The stream is divided into 8 segments, from $\phi_1$ = $-4^\circ$ to 4$^\circ$, spaced by 1$^\circ$. For dispersions of the model, we directly calculate the standard deviations of observables in each slice. For the observations, we describe observables with a Gaussian distribution by taking into account the observed uncertainties on individual stars, except for $\phi_2$ where the uncertainty is nearly zero. A uniform prior is used for all dispersions. The posteriors of dispersions are derived using MCMC sampling. Their 16th, 50th, and 84th percentiles are shown in Figure~\ref{fig:model_vs_data}. Note that the actual progenitor of the stream is not detected as it is probably completely disrupted, and thus we refer to it as a postulated progenitor.

Overall, we find good agreement between the simulation and observation. As shown by the sky position panel (Figure~\ref{fig:model_vs_data}a), the spur ($\phi_1$, $\phi_2$) $\sim$ ($-3$, 0.5)$^\circ$ and fanning ($\phi_1$ $>$ 1$^\circ$) features are well reproduced. Although the spatial distribution of model particles does not perfectly match the observed distribution at the far left edge of the stream, the stream length is still short enough to be comparable to what is observed given its long evolution time of 3 Gyr in the simulation. In both panels of $\mu^*_\alpha$ and $V_r$, (Figure~\ref{fig:model_vs_data}c and Figure~\ref{fig:model_vs_data}g), the model seemingly exhibits a single sequence along $\phi_1$, whereas for $\mu_\delta$ (Figure~\ref{fig:model_vs_data}e), the stream is much more chaotic with multiple tributaries overlapping (see also Figure~\ref{fig:bar_const_axi}b). The other panels display dispersions (or widths) of each observable for both data and the model, where we also note compatible trends.

\subsection{Dissecting the model particles} \label{subsec:groups}
\begin{figure}[htb!]
\centering
\includegraphics[width=\columnwidth]{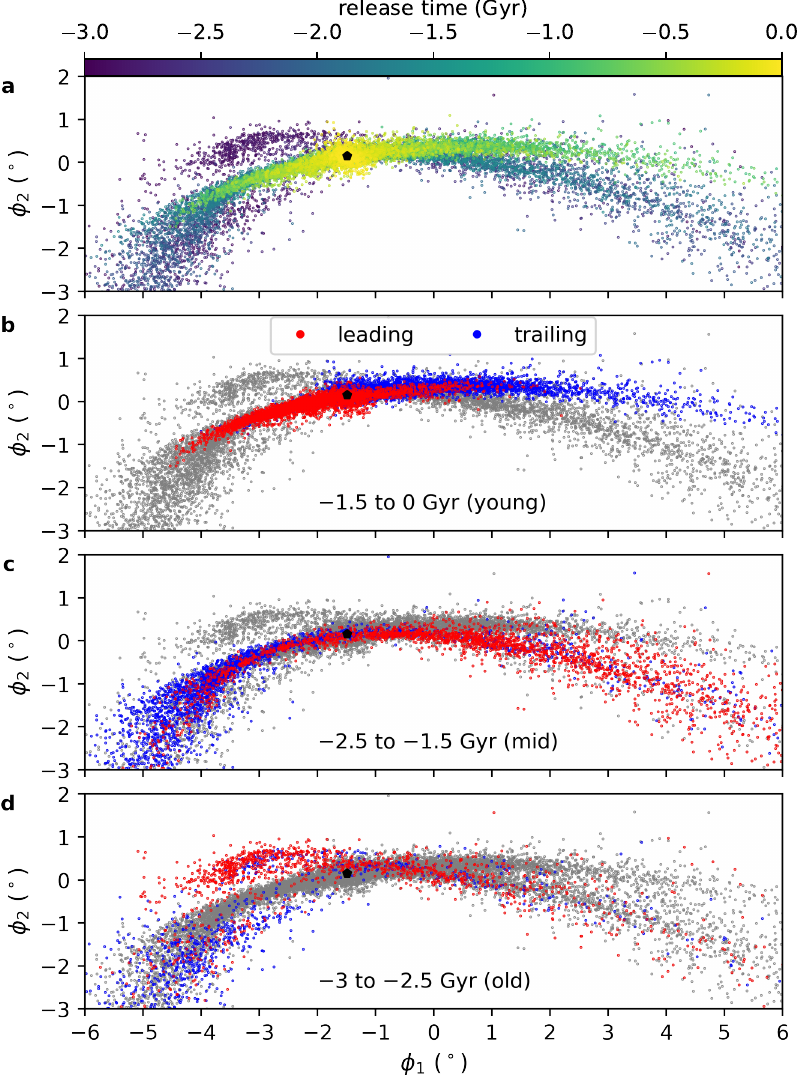}
\caption{Dissection of the model stream. The black pentagon represents the postulated progenitor. \textbf{a}, model particles color coded by release time. The remaining three panels show particles released within different time slices: from $-1.5$ Gyr ago to now (\textbf{b}), $-2.5$ to $-1.5$ Gyr (\textbf{c}), and $-3$ to $-2.5$ Gyr (\textbf{d}), respectively. We further classify particles as leading or trailing based on their energy relative to the progenitor at their respective release time. The grey dots in \textbf{b}-\textbf{d} indicate all model particles.}
\label{fig:slice}
\end{figure}

We record the release time of all model particles when they are detached from the progenitor. Figure~\ref{fig:slice}a displays the model of Ophiuchus with particles color-coded by release time. In general, it is expected that particles disrupted earlier should be found farther away from the progenitor and the most recently released particles should be concentrated near the progenitor (see Ophiuchus under an axisymmetric potential in Figure~\ref{fig:bar_const_axi}). The latter is verified in Figure~\ref{fig:slice}a, but our model does not display the former behavior. Instead, we see that the particles released earliest (i.e. the ``oldest" particles) form the spur, while particles released more recently (i.e. ``younger" particles) act to form the fanning features.

Based on the morphology of the stream, we divide all particles by their release time into three groups: $-1.5$ to 0 Gyr (``young" group, where we define 0 as the present time), $-2.5$ to $-1.5$ Gyr (``mid" group), and $-3$ to $-2.5$ Gyr (``old" group). We also compute the total energy (potential plus kinetic energy) of each particle at its corresponding release time, and classify it into the leading (trailing) tail if its energy is smaller (greater) than that of the progenitor. One thing to be clarified is that stars with lower (higher) energy usually orbit the Galaxy faster (slower), which makes them leading (trailing). Figures~\ref{fig:slice}b-d presents these three groups with leading and trailing arms labeled. With this labeling, each feature in the model appears clearer and we can distinguish in more depth that the leading old group accounts for the spur, while the fanning is made of the trailing young plus leading mid groups. 

\begin{figure*}[htb!]
\centering
\includegraphics[width=\textwidth]{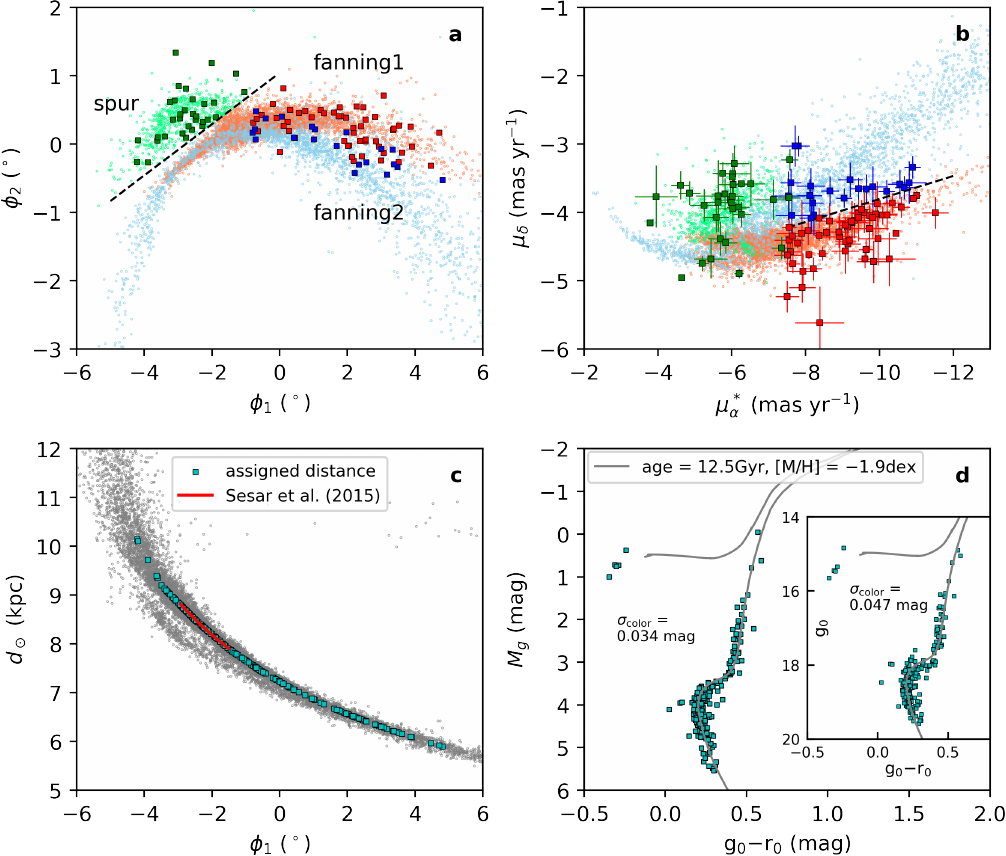}
\caption{Cross-validation between the Ophiuchus model and data. In \textbf{a} and \textbf{b}, the spur, young trailing group (which we refer to as fanning1), and mid leading group (referred to as fanning2) of the model are plotted in light green, light red, and light blue dots respectively. Member stars in data that we believe correspond to these features are presented in green, red, and blue squares, respectively. The black dashed lines show how we define the spur (\textbf{a}), fanning1 and fanning2 (\textbf{b}). \textbf{c}, distances assigned to the data (cyan) based on the model (gray). The distance gradient from \citet{2015ApJ...809...59S} is shown in the red line, which matches our model. \textbf{d}, on a color-magnitude diagram, we compare member stars to a stellar track of (age, [M/H]) = (12.5 Gyr, $-1.9$ dex) after converting apparent magnitudes of stars into the absolute magnitudes by using distances in \textbf{c}. The subfigure shows the same but without correcting for individual distances, i.e., apparent magnitudes, where the isochrone is at the stream's mean distance of 7.9 kpc \citep{2022ApJ...928...30L}. Here magnitudes in $g-$ and $r-$ band are photometry taken from the Dark Energy Camera \citep{2015AJ....150..150F,2019AJ....157..168D,2021ApJS..255...20A,2022ApJS..261...38D},  dereddened based on \citet{1998ApJ...500..525S,2011ApJ...737..103S} using \texttt{dustmaps} python package \citep{2018JOSS....3..695G}.}
\label{fig:validate}
\end{figure*}

We further validate our model by investigating the spur and fanning in both the data and model, which is illustrated in Figure~\ref{fig:validate}a-b. Here the spur is defined as those points whose positions are above the black dashed line in Figure~\ref{fig:validate}a, which is given by
\begin{equation}
\phi_2 = 0.375 \phi_1 + 1.041 .
\end{equation}
From Figures~\ref{fig:validate}a to \ref{fig:validate}b, it is noticeable that the model predicts the spur's proper motions well. Given that the fanning consists of two parts in our model, we further divide it into ``fanning1'' (red for data and light red for model) and ``fanning2'' (blue and light blue for data and model, respectively) based on $\mu_\delta$ using the black dashed line in Figure~\ref{fig:validate}b defined by
\begin{equation}
\mu_\delta = -0.168 \mu^*_\alpha - 5.485 
\; (\mu^*_\alpha < -7.5 \; {\rm mas \; yr^{-1}}) ,
\end{equation}
since they are more distinguishable in $\mu_\delta$. Looking from Figure~\ref{fig:validate}b back to \ref{fig:validate}a, we find that the observed stars that belong to fanning1 according to their proper motions are located at higher $\phi_2$ than the fanning2 stars, as predicted by the model. Here we do not inspect the radial velocity because the spur can hardly be distinguished from the main stream track and the two fanning features also overlap, both in the model and data.

Due to the absence of accurate distances in the data, we assign each star a distance according to the model particles given its $\phi_1$, and this is illustrated in Figure~\ref{fig:validate}c. The distance gradient measured from \citet{2015ApJ...809...59S} is also shown, which matches with our model. By adjusting all stars to their absolute magnitudes using assigned distances, we can obtain a clear stellar track following an isochrone \citep[extracted from the PARSEC database,][]{2012MNRAS.427..127B} on a color-magnitude diagram in Figure~\ref{fig:validate}d, where the small dispersion in color corroborates the globular cluster origin for the stream as pointed out by \citet{2022ApJ...928...30L}, based on the stream's low velocity and metallicity dispersions. Here we quantify the color dispersion by
\begin{equation}
\sigma_\mathrm{color} = \sqrt{ \frac{1}{N} \sum_i^N 
( \mathrm{color_{iso}}(\mathrm{gmag}_i) - \mathrm{color}_i  )^2 }   ,
\end{equation}
where $\mathrm{color}_i$ is a star's color and $\mathrm{color_{iso}}(\mathrm{gmag}_i)$ stands for the color derived from the isochrone at given $g$ magnitude of the star. This calculation only applies to the main sequence and red giant branch stars with color $>$ 0 mag, because color is no longer a function of $g$ for the horizontal branch. A smaller $\sigma_\mathrm{color}$ = 0.034 mag is obtained. As a comparison, the situation in which individual distances are not corrected is shown in the subfigure (the isochrone is at Ophiuchus' mean distance of 7.9 kpc  \citep{2022ApJ...928...30L}), where $\sigma_\mathrm{color}$ = 0.047 mag means that stars do not match the isochrone well. This also indicates that the prediction of distance from our model is valid.

\subsection{Flipping of leading and trailing} \label{subsec:snapshot}
From Figure~\ref{fig:slice}, it can be seen that the model's leading and trailing tails have complicated orientations for the old and mid groups when compared to the young group. Normally in a stellar stream, the leading tail is the component of the tidal tails that travels in front of the progenitor while the trailing tail follows behind, and two tidal components usually stay in this arrangement as they continue their orbits. Ophiuchus is currently orbiting from positive $\phi_1$ to negative $\phi_1$, meaning that negative and positive $\phi_1$ (i.e. the left and right sides of figures plotted against $\phi_1$) should correspond to the leading and trailing tails, respectively. This is verified by the young group in Figure~\ref{fig:slice}b. However, the mid group in Figure~\ref{fig:slice}c shows the completely opposite situation, where the ``leading tail'' now orbits behind the progenitor and the ``trailing tail'' orbits in front of the progenitor. In addition, the old group of Figure~\ref{fig:slice}d has two tidal tails both mostly located on the leading side. We attribute this to torques from the Galactic bar when Ophiuchus repeatedly comes close to its pericenter, and particles released earlier ($<$ $-1.5$ Gyr) are torqued more strongly. The detailed mechanism is elaborated on in Section~\ref{subsec:torque}. 

\begin{figure}[htb!]
\centering
\includegraphics[width=\columnwidth]{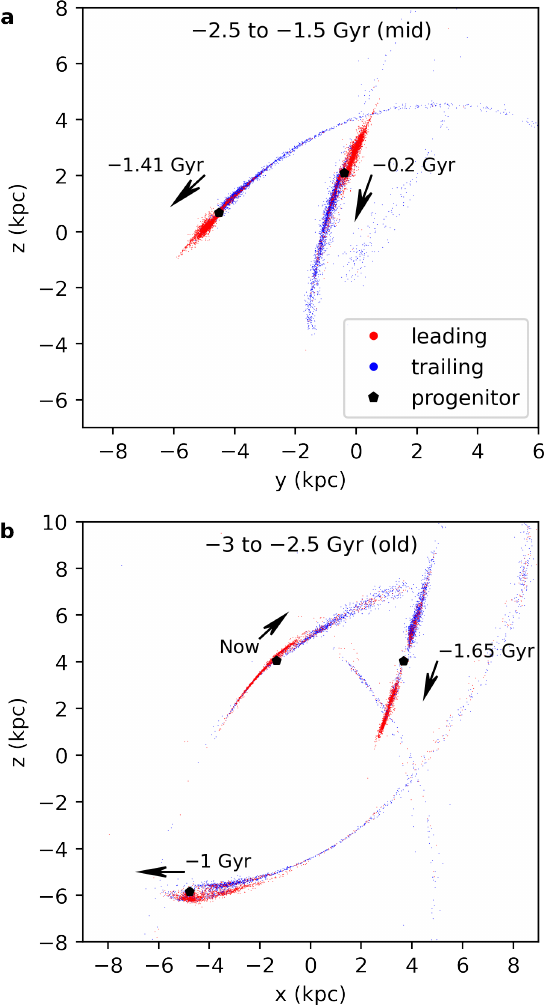}
\caption{Snapshots of the mid and old groups at several time points. \textbf{a}, the assumed progenitor, leading and trailing tails of the mid group, are plotted in the black pentagon, red and blue dots respectively in the Galactocentric Cartesian ($y$, $z$) plane. We show two moments in time ($-1.41$ and $-0.2$ Gyr) before and after the spatial flipping of the tails. The black arrows indicate their moving directions. \textbf{b}, the same but for the old group in the ($x$, $z$) plane.}
\label{fig:snapshot}
\end{figure}

To show the flipping and transposition clearly, we present several snapshots for the mid and old groups in Figure~\ref{fig:snapshot}. From Panel \ref{fig:snapshot}a, the mid group's leading and trailing tails are distributed in the expected positions at, for example, $-1.41$ Gyr, but they completely flip at $-0.2$ Gyr. Similarly, Panel \ref{fig:snapshot}b displays snapshots of the old group, the morphology of which is more diverse and complex. At $-1.65$ Gyr, the postulated progenitor is in the middle with the leading tail ahead and the trailing tail behind. At $-1$ Gyr, the leading tail is overtaken by the progenitor and is orbiting alongside the trailing tail. At the present time, the trailing tail is ahead of the progenitor position while the leading tail lags behind it.

Furthermore, it can be recognized that a small amount of particles fly away apart from the main stream body in Figure~\ref{fig:snapshot} (see also Figure~\ref{fig:validate}c), because they are dispersed by the bar into chaotic orbits forming underdensities, as pointed out by \citet{2016ApJ...824..104P}. However, the fraction of them only takes up $\sim$ 5\% and 95\% of model particles are concentrated around the postulated progenitor.

\subsection{The Galactic bar is torquing} \label{subsec:torque}
\begin{figure*}[htb!]
\centering
\includegraphics[width=\textwidth]{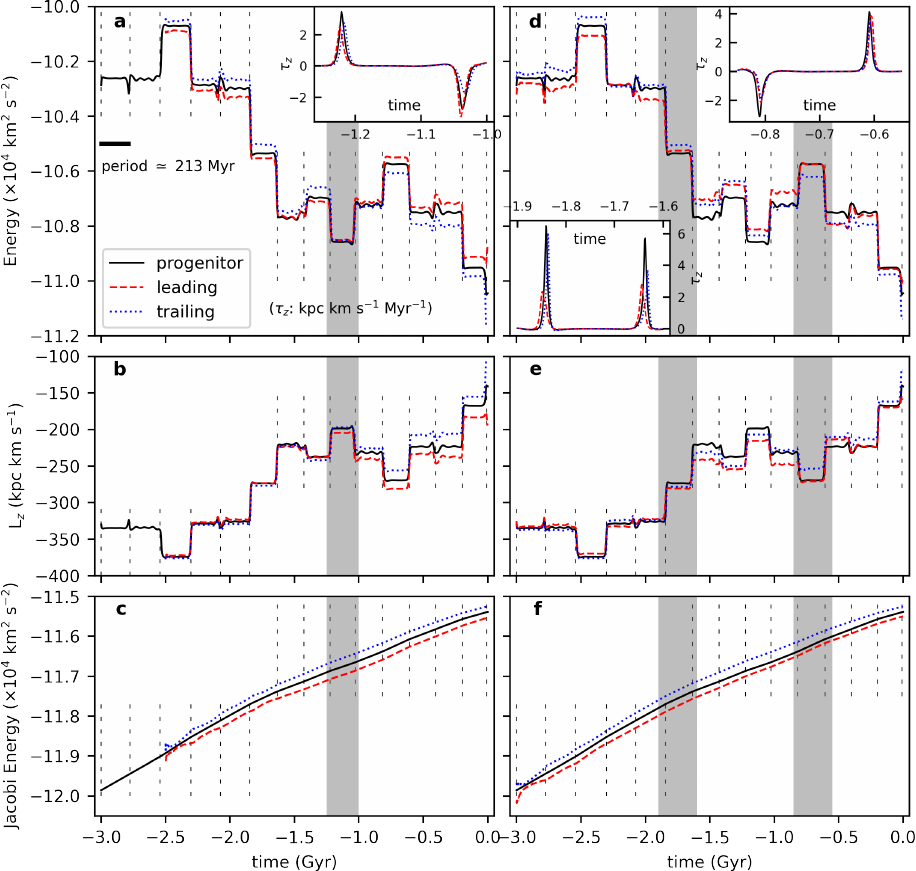}
\caption{Evolution of specific (per unit mass) energy, z-component torque and angular momentum, as well as Jacobi energy. \textbf{a}, energy as a function of time for the progenitor (black solid), leading (red dashed) and trailing (blue dotted) tails of the mid group. The subplot shows their z-component torques (in units of kpc km s$^{-1}$ Myr$^{-1}$) during a specified time interval (gray area in the main panel). The mean orbital period over the past 3 Gyr is $\sim$ 213 Myr. \textbf{b}, z-component angular momentum of the mid group. \textbf{c}, Jacobi energy of the mid group. \textbf{d}-\textbf{f}, the same but for the old group. Here we show the specific torques of two indicated time intervals in subpanels. Vertical dashed lines in all panels indicate pericenter passages of the postulated progenitor.}
\label{fig:history}
\end{figure*}

Figure~\ref{fig:snapshot} implies a complicated orbital history for the leading and trailing tails of the mid and old groups. To reveal what causes these transpositions, in Figure~\ref{fig:history} we inspect their dynamical evolution, including energy ($E$), z-component torque ($\tau_z$) and angular momentum ($L_z$), as well as Jacobi energy equal to $E_J$ = $E$ $-$ $\Omega \cdot L_z$, where $\Omega$ is the bar's pattern speed. 

As stated in Section~\ref{subsec:model}, the modeling process spans 3 Gyr with 1 Myr as the time step (i.e. 3,000 steps in total). At each step, for particles that have been released by that point belonging to either the leading or trailing tails of the mid or old groups, we compute median values of their energy, $\tau_z$, $L_z$, and $E_J$. By doing so, we obtain corresponding changes as a function of time as shown in Figure~\ref{fig:history}. Figures~\ref{fig:history}a-c show the case of the mid group while Figures~\ref{fig:history}d-f show the old group. From Figures~\ref{fig:history}a,b,d,e, the stream's energy, $\tau_z$, and $L_z$ are repeatedly increased or decreased at pericenter passages due to the bar. When the stream is further from the Galactic center, the energy and $L_z$ are conserved and the torque falls back to zero because the outer Galactic potential is nearly axisymmetric.

In Figure~\ref{fig:history}a, we show the leading and trailing tails of the mid group (starting from $-2.5$ Gyr) compared to the assumed progenitor. The leading (trailing) tail has lower (higher) energy when initially released from the progenitor. This discrepancy continues until two pericenters near $-1$ Gyr, marked by the gray region, after which their signs of energy flip with respect to the progenitor. We present the corresponding $\tau_z$ during this time interval in the subpanel. Keeping in mind that a right-handed Cartesian coordinate is used such that $L_z$ of both the bar and Ophiuchus are negative (Figure~\ref{fig:history}b, though the stream's orbit is more polar than planar, see Figure~\ref{fig:pot}b), positive $\tau_z$ represents a ``pulling-backward" force that decreases energy while negative $\tau_z$ represents a ``pushing-forward" force that increases energy. Although the sign of $\tau_z$ is the same at the two pericenters, the magnitudes of the torque imposed on the leading and trailing tails are different, which causes the aforementioned energy flip, and subsequent spatial flip of the two tails after some time. We also point out that during this process, a pulling force trying to slow down a tail (here positive $\tau_z$) will make stars fall in towards the Galactic center, and potential energy will transform into kinetic energy, which results in the stars moving faster. In contrast, a thrusting force (here negative $\tau_z$) will cause stars to move slower instead (see Section~\ref{subsec:explain} for an elaboration). This is why the original trailing tail can overtake the leading tail.

From Figure~\ref{fig:history}b, $L_z$ shows an opposite-direction evolution with respect to energy, and we can still identify similar flipping around the two pericenters marked by the gray region. For $E_J$ in Figure~\ref{fig:history}c, it evolves fairly smoothly without the sudden rise or drop that happens in energy and $L_z$, nor is there any flipping. This is because energy and $L_z$ are not conserved in such a time-dependent barred potential, while $E_J$ is nearly conserved during each pericenter passage since the pattern speed does not change much over the timescale of the pericenter. Given $E_J$ = $E$ $-$ $\Omega \cdot L_z$ $\simeq$ constant, we can also infer that $E$ and $L_z$ evolve conversely considering $- \Omega$ $>$ 0 for the Milky Way. However, as the bar keeps decelerating, $E_J$ cannot remain conserved over the whole 3 Gyr and changes by half as much as the energy.

Figures~\ref{fig:history}d-f display results for the old group from $-3$ Gyr onward. We note two pairs of pericenters in Figure~\ref{fig:history}d that rearrange the energy of both tails onto one side relative to the progenitor. Two pericenters between $-2$ and $-1.5$ Gyr let the stream experience two successive positive $\tau_z$. The applied torques are comparable on the progenitor and trailing tail, but significantly less severe on the leading tail, placing the energy of both tails above that of the progenitor. After some time, we can see that the progenitor and trailing tail begin to overtake the leading tail at $-1$ Gyr in Figure~\ref{fig:snapshot}b. Note that it can take some time for the spatial transposition to happen after the energy flip. As for the two pericenters between $-1$ to $-0.5$ Gyr, the overall effects of $\tau_z$ put the energy of both tidal tails below that of the progenitor, meaning the progenitor starts to lag behind the tails. Moreover, we can also deduce that the progenitor is ahead of both tails during this period because its $\tau_z$ peak comes first, verifying the snapshot at $-1$ Gyr in Figure~\ref{fig:snapshot}b. Figure~\ref{fig:history}e presents $L_z$ of the old group, from which we note similar reallocations between both tails and the progenitor within two time intervals (gray area). There seems to be a $L_z$ flip at $-2.3$ Gyr that, however, does not necessarily indicate a real flipping of tidal tails because the key lies in energy rather than $L_z$. Finally, Figure~\ref{fig:history}f displays a similar smooth evolution of $E_J$.

\begin{figure}[htb!]
\centering
\includegraphics[width=\columnwidth]{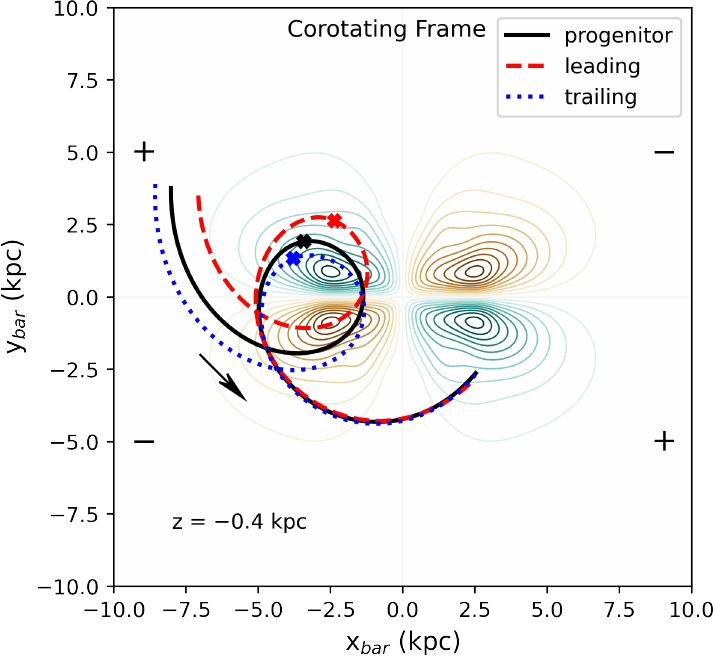}
\caption{Illustration of torques around a pericenter. The black, red and blue lines show orbits of the progenitor, the leading and trailing tails of the old group respectively, between $-1.9$ and $-1.75$ Gyr in ($x$, $y$) plane corotating with the bar. Three ``X'' indicate where maximum $\tau_z$ reaches. Cyan and yellow contours indicate positive and negative $\tau_z$ at $z$ = $-0.4$ kpc, the plane near where the maximum $\tau_z$ occurs. The contour has 21 levels ranging from $-$10 to 10 spacing by 1 kpc km s$^{-1}$ Myr$^{-1}$. }
\label{fig:torque}
\end{figure}

Figure~\ref{fig:torque} illustrates the orbits of the old group around the time of the pericenter between $-1.9$ to $-1.75$ Gyr in the ($x$, $y$) plane corotating with the bar. Cyan and yellow contours indicate positive and negative $\tau_z$ at $z$ = $-0.4$ kpc, the plane near where the maximum $\tau_z$ occurs. We see directly that the torque felt by the progenitor and trailing tail is higher (i.e. a more positive $\tau_z$)  than that of the leading tail (the ``X'' marks where the maximum $\tau_z$ comes), corresponding to the first peak in the left-bottom subpanel of Figure~\ref{fig:history}d.

\subsection{Dragging a star makes it faster instead of slower and vice versa} \label{subsec:explain}
To understand why the tidal tails of Ophiuchus can flip or transpose, we can analyze an illustrative example of a simple circular motion. Imagine that there is a particle with mass $m$ circularly orbiting a massive object in the center with mass $M$. The radius and velocity scalar of the particle are initially $r_1$ and $v_1$. It is the gravity that supports the motion, and we have
\begin{equation}
\frac{GMm}{r^2_1} = m \frac{v^2_1}{r_1}.
\label{eq1}
\end{equation}
Now let $v_1$ change by $\Delta_v$ suddenly to imitate the effects of torques from the bar, which breaks the balance in Equation~\ref{eq1}. We assume that after achieving balance again, the particle still closely follows a circular orbit with a new radius and velocity of $r_2$ and $v_2$. Similarly, there is
\begin{equation}
\frac{GMm}{r^2_2} = m \frac{v^2_2}{r_2}.
\label{eq2}
\end{equation}
Because energy is conserved in this system after the interaction, we have
\begin{equation}
\frac{1}{2} m (v_1 + \Delta_v)^2 - \frac{GMm}{r_1} = \frac{1}{2} m v^2_2 - \frac{GMm}{r_2}.
\end{equation}
Substituting terms of potential with Equation~\ref{eq1} and \ref{eq2}, we can obtain
\begin{equation}
v^2_1 - 2 v_1 \Delta_v - \Delta^2_v = v^2_2 .
\end{equation}
If $\Delta_v$ $>$ 0, which represents the increase of energy (this mimics negative $\tau_z$ in the Ophiuchus case), it is true that $v_1$ $>$ $v_2$, meaning that the particle is slower after the interaction. If $\Delta_v$ $<$ 0, corresponding to decreased energy (positive $\tau_z$), we have $-2v_1\Delta_v$ $-$ $\Delta^2_v$ $>$ 0 given $-\Delta_v$ $<$ $v_1$, and then $v_1$ $<$ $v_2$, because if $-\Delta_v$ $>$ $v_1$ the particle will turn back, which does not happen in Ophiuchus. Based on this simplified case, we can conclude that a force trying to accelerate stars will slow them down, and a stopping force will make them faster instead.

We can reach the same result with a more general treatment by analyzing the following equation from \citet{2015MNRAS.450.1136E},
\begin{equation}
\Delta\varphi(t) = - A \Delta^\parallel_v t + 
    B \frac{\Delta^\parallel_v}{v} {\rm sin}(Cvt) - 
    D \frac{\Delta^\perp_v}{v} \left[ 1- {\rm cos}(Cvt) \right] ,
\end{equation}
which describes how particles on circular orbits in a spherical potential would respond after a velocity kick. Here $A$, $B$, $C$, and $D$ are some constant coefficients where $A$ $>$ 0. $v$ is the stream velocity before the kick while $\Delta^\parallel_v$ and $\Delta^\perp_v$ are velocity changes parallel and perpendicular to the stream, respectively. $\Delta\varphi(t)$ represents the angular offset of nearby particles relative to the kick location. To simplify the analysis, we can assume that there is only velocity kick along the stream, i.e., $\Delta^\perp_v$ = 0. The second part of the equation is an oscillating term that only matters shortly after the kick. After sufficient time, only the first term is dominant, meaning that $\Delta\varphi$ changes inversely to $\Delta^\parallel_v$. Thus, a positive kick causes particles to lag, while a negative kick results in overtaking, the same as above.

Having the aforementioned conclusion, we can look again at the subpanel of Figure~\ref{fig:history}a. Both the leading and trailing tails of the mid group first experience positive $\tau_z$ meaning ``pulling-backward'' forces which will speed them up. Whereas, stronger $\tau_z$ accelerates the trailing more than the leading with weaker $\tau_z$. When the second pericenter comes, negative $\tau_z$ acts as ``pushing-forward'' forces that eventually slow down both tails. However, the trailing is decelerated less than the leading because $\tau_z$ magnitude of the former is smaller than that of the latter. Therefore, the trailing overtakes the leading and there is the flip. The transposition between the postulated progenitor and both tails of the old group can be interpreted in a similar way.

\subsection{Exploring Ophiuchus under various rotations of the bar} \label{subsec:bars}
The above model is built upon the fiducial barred Milky Way potential where a uniform deceleration is adopted with current pattern speed $\Omega_0$ = $-$35 km s$^{-1}$ kpc$^{-1}$ and slowing rate $a$ = 5.5 km s$^{-1}$ kpc$^{-1}$ Gyr$^{-1}$. In the following, we explore the Ophiuchus in a variety of Galactic potentials.

\begin{figure}[htb!]
\centering
\includegraphics[width=\columnwidth]{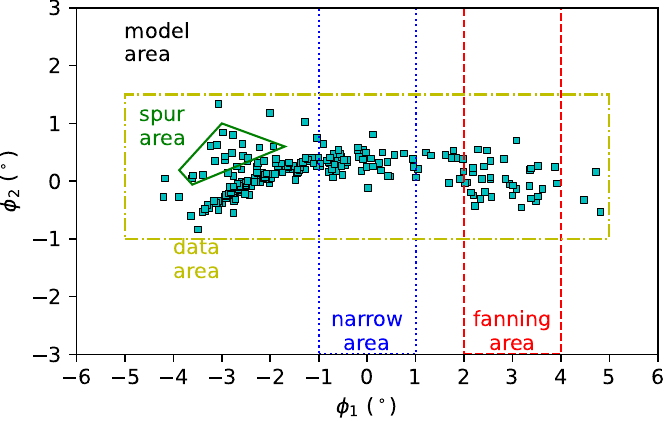}
\caption{Definitions of various areas used in Equation~\ref{eq:lns} based on the data. The spur area is a polygon enclosed by vertices ($\phi_1$, $\phi_2$) = ($-3.61$, $-0.06$), ($-1.70$, 0.60), ($-3.00$, 1.00) and ($-3.88$, 0.19). The rest areas are all rectangles: the data area from ($-5.0$, $-1.0$) to (5.0, 1.5), the narrow area from ($-1.0$, $-3.0$) to (1.0, 3.0), the fanning area from (2.0, $-3.0$) to (4.0, 3.0), and the model area from ($-6.0$, $-3.0$) to (6.0, 3.0).}
\label{fig:area_def}
\end{figure}

Prior to beginning, we customize a grading criteria to quantitatively estimate how close a model stream's morphology is to the data. As mentioned above, Ophiuchus has unusual features including the spur, fanning, and short length. By taking these into account, we define the score 
\begin{equation}
\mathrm{ln}\mathcal{S} = \mathrm{ln}f_\mathrm{sa} + 
    \mathrm{ln}f_\mathrm{da} + \mathrm{ln}\frac{\sigma_\mathrm{fa}}{\sigma_\mathrm{na}} - \chi^2_\mathrm{ma} .
\label{eq:lns}
\end{equation}
$f_\mathrm{sa}$ is the fraction of model particles populating in the ``spur area'' in Figure~\ref{fig:area_def}, and a higher $f_\mathrm{sa}$ means a stronger spur. Similarly, $f_\mathrm{da}$ is the fraction of particles within the ``data area'', and a bigger $f_\mathrm{da}$ implies that a model stream has a shorter length because fewer particles are outside of this area. $\sigma_\mathrm{fa}$ and $\sigma_\mathrm{na}$ stand for $\phi_2$ dispersions of particles located in the ``fanning area'' and ``narrow area'', respectively, and their ratio measures how spread a fanning is. $\chi^2_\mathrm{ma}$ is given by 
\begin{equation}
\chi^2_\mathrm{ma} = \frac{1}{N_\mathrm{bin}} \sum^{N_\mathrm{bin}}_{i=1}
    \frac{(\phi_\mathrm{2,model}-\phi_\mathrm{2,data})^2}
    {\sigma^2_\mathrm{\phi_2,model}+\sigma^2_\mathrm{\phi_2,data}}
\end{equation}
where $N_\mathrm{bin}$ = 8 ($\phi_1$ from $-4^\circ$ to $4^\circ$ spacing $1^\circ$), $\phi_\mathrm{2,model}$ and $\phi_\mathrm{2,data}$ represent median $\phi_2$ locations of a model within the ``model area'' and the data in each $\phi_1$ bin, $\sigma_\mathrm{\phi_2,model}$ and $\sigma_\mathrm{\phi_2,data}$ are corresponding $\phi_2$ dispersions. Thus $-\chi^2_\mathrm{ma}$ measures consistency of trajectories between a model and the data. Here $\sigma_\mathrm{\phi_2,data}$ are values in Figure~\ref{fig:model_vs_data}b, and for $\sigma_\mathrm{fa}$, $\sigma_\mathrm{na}$, and $\sigma_\mathrm{\phi_2,model}$ we use the half difference of 84th and 16th percentile to cope with complicated chaotic morphologies of various models.  

\begin{figure*}[htb!]
\centering
\includegraphics[width=\textwidth]{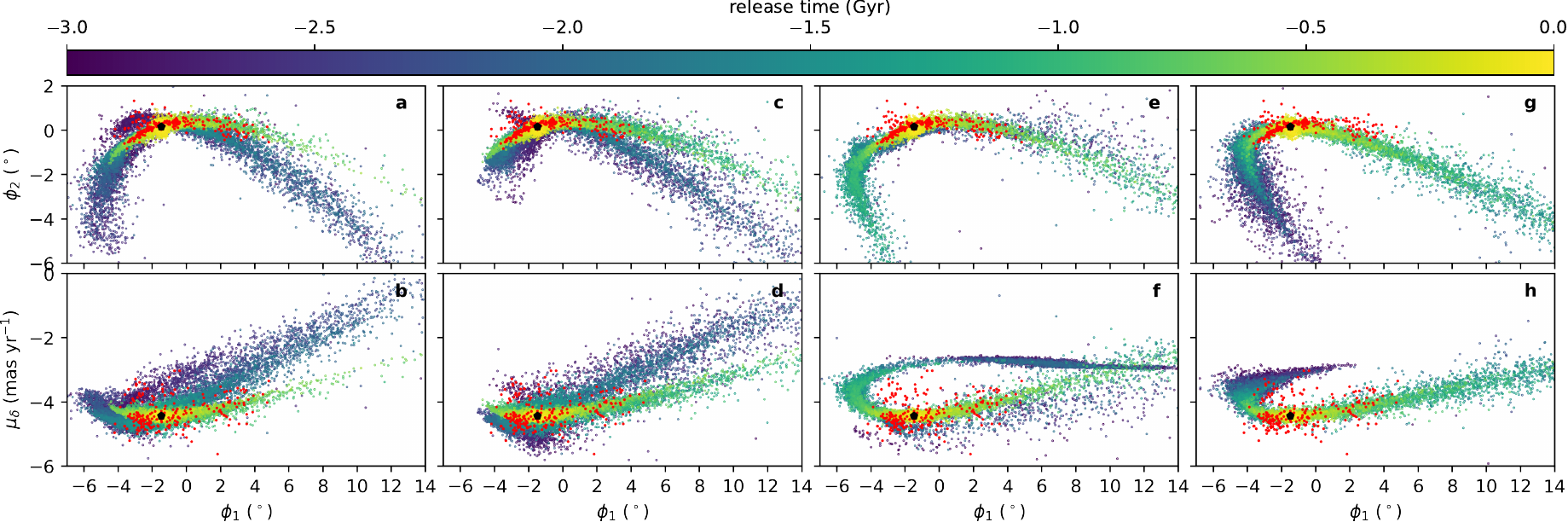}
\caption{Comparison of Ophiuchus model streams under different potentials. \textbf{a,b}: $\phi_2$ and $\mu_\delta$ vs. $\phi_1$ of the fiducial model stream within the barred potential analyzed in this work. Particles are color-coded by the release time. \textbf{c,d}: similar to the previous, but modeled with a bar rotating following Equation~\ref{eq:chiba}. \textbf{e,f}: a model stream with a constant pattern speed $\Omega$ = $-$35 km s$^{-1}$ kpc$^{-1}$. \textbf{g,h}: a model stream obtained under an axisymmetric potential with the similar mass profile to the barred one. A portion of the trailing tails of the latter two is truncated due to the long length. The Ophiuchus stars are overplotted in red dots.}
\label{fig:bar_const_axi}
\end{figure*}

Firstly, we inspect differences in modeling Ophiuchus when employing potentials with a non-uniformly decelerating bar, a constantly rotating bar or without a bar at all; these are shown in Figure~\ref{fig:bar_const_axi}c-h, while Figure~\ref{fig:bar_const_axi}a,b display the fiducial model as comparison. In Figure~\ref{fig:bar_const_axi}c,d, we test the mode of \citet{2021MNRAS.500.4710C} as shown in Figure~\ref{fig:pot}c, therein $\eta$ = $\dot{\Omega}/\Omega^2$ leads to $\Omega$ = $\Omega_0 / (1 - \eta \Omega_0 t)$ further giving 
\begin{equation}
\theta = \theta_0 - \frac{1}{\eta} \ln (1 - \eta \Omega_0 t)
\label{eq:chiba}
\end{equation}
\citep[Note that the Cartesian $z$ direction used here is opposite in][]{2021MNRAS.500.4710C}. We let $\eta$ = 0.003 \citep{2021MNRAS.500.4710C,2023MNRAS.525.3576C} with other parameters the same as the fiducial ones. From Figure~\ref{fig:bar_const_axi}e,f to g,h are model streams generated under a barred potential with fixed $\Omega$ = $-$35 km s$^{-1}$ kpc$^{-1}$ and generated within an axisymmetric potential. Since these models all have similar distributions in $\mu^*_\alpha$ and $V_r$ to the fiducial one as presented in Figure~\ref{fig:model_vs_data}c,g, Figure~\ref{fig:bar_const_axi} only lists their $\phi_2$ and $\mu_\delta$ along $\phi_1$, which are more distinguishable from each other. The Ophiuchus stars are overplotted in red dots to enhance the contrast. 

Figure~\ref{fig:bar_const_axi}a presents the same morphology of the fiducial model as Figure~\ref{fig:slice}a but in a wider context. From panel b, it can be noticed that the young, mid, and old groups are arranged in an orderly overlapping manner in $\mu_\delta$ after being color-coded by release time. The second model in panel c,d resembles the first one owning a spur and fanning as well because the bar's pattern speed in two models evolves similarly (Figure~\ref{fig:pot}c). However, its spur is far less prominent, and its ln$\mathcal{S}$ is scored only $-4.9$, compared to the fiducial ln$\mathcal{S}$ = $-3.064$. In addition, it is clear that the rest two (panel e-h) fail to recover the short length, spur, and fanning of the Ophiuchus, and do not possess multiple overlapping tributaries as seen in the fiducial model. Apparently, the steady bar (panel e,f) has dispersed more ``old'' particles (darker) beyond the scope inspected here, implying that there is a longer stream, and its ln$\mathcal{S}$ = $-9.911$. For the axisymmetric case (panel g,h), the stream undergoes a ``normal'' unperturbed evolution and particles released earlier populate farther from the postulated progenitor just as expected, which contrasts with the fiducial one. It also spans a much longer angular extent (truncated here) and has a ln$\mathcal{S}$ of $-7.708$. 

\begin{figure*}[htb!]
\centering
\includegraphics[width=\textwidth]{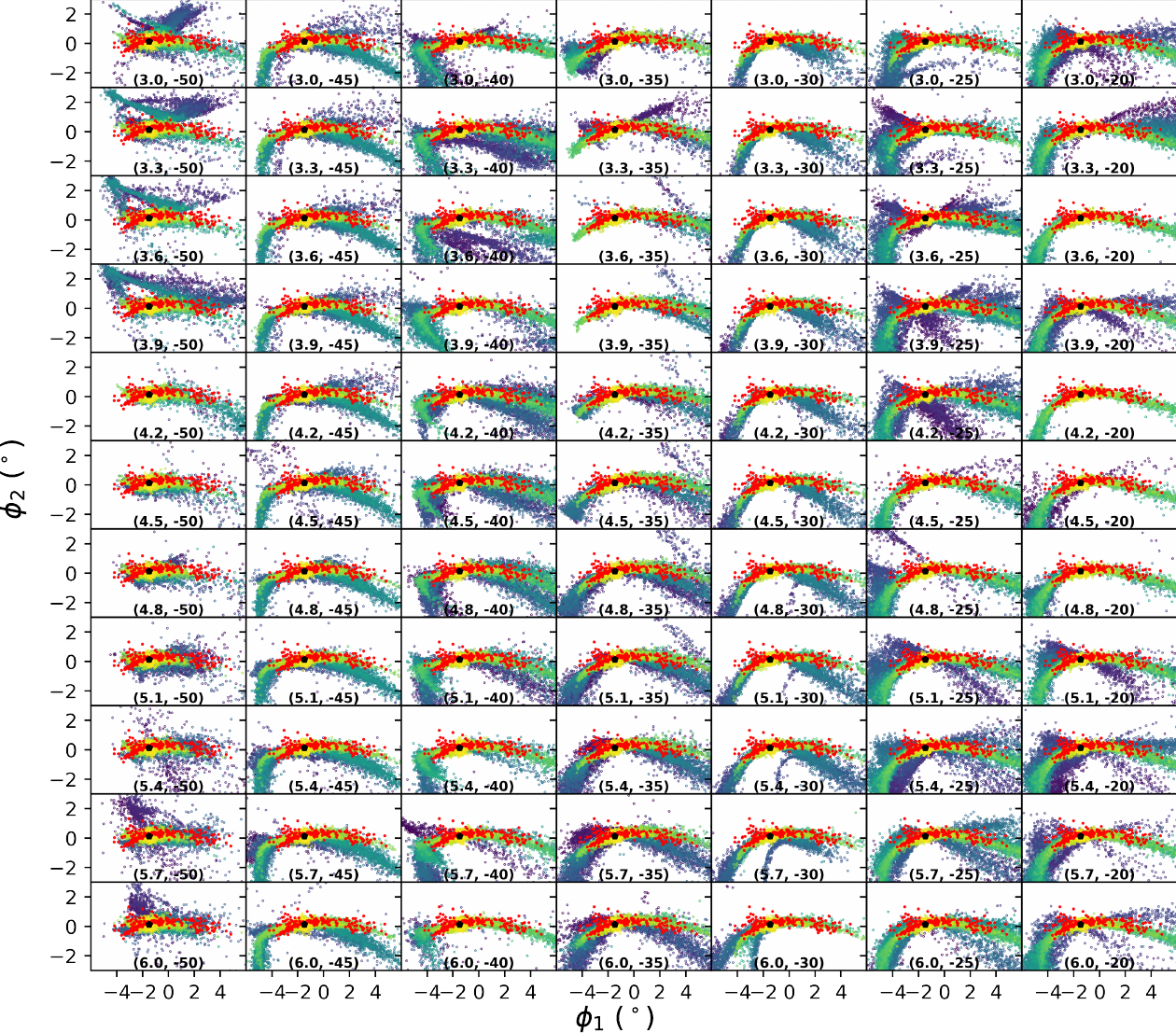}
\caption{Morphologies of the Ophiuchus stream modeled over a grid of $a$ and $\Omega_0$ in Equation~\ref{eq:uniform}. From top to bottom, $a$ changes from 3.0 to 6.0 with spacing 0.3 km s$^{-1}$ kpc$^{-1}$ Gyr$^{-1}$. From left to right, $\Omega_0$ varies from $-$50 to $-$20 spaced by 5 km s$^{-1}$ kpc$^{-1}$. The colormap is the same as in Figure~\ref{fig:bar_const_axi}. Red dots show our $S^5$ data.}
\label{fig:explore_bar}
\end{figure*}

Additionally, we explore the Ophiuchus response to the parameters in Equation~\ref{eq:uniform}. Instead of ``static'' $\theta_0$, we are more interested in its ``dynamical'' derivatives $\Omega_0$ and $a$. Specifically here, $\theta_0$ is fixed as $-30^\circ$, and $\Omega_0$ is varied from $-$50 down to $-$20 km s$^{-1}$ kpc$^{-1}$ with a spacing of 5 km s$^{-1}$ kpc$^{-1}$, while $a$ spans from 3.0 to 6.0 km s$^{-1}$ kpc$^{-1}$ Gyr$^{-1}$ spaced by 0.3 km s$^{-1}$ kpc$^{-1}$ Gyr$^{-1}$. The stream is modeled over the two-dimensional grid and its morphology in ($\phi_1$, $\phi_2$) is displayed in Figure~\ref{fig:explore_bar} accordingly. The numbers in each plot indicate $a$ and $\Omega_0$, respectively. The colors denote the same as Figure~\ref{fig:bar_const_axi}. 

Overall, the Ophiuchus stream is heavily disturbed due to the bar's rotation \citep{2016ApJ...824..104P}, showing complicated extended structures. Some of them can account for the short length, e.g. $\Omega_0$ = $-$50 km s$^{-1}$ kpc$^{-1}$; some display various spur features, e.g. (3.3, $-$35) or (3.9, $-$20); the fanning exists under many situations, e.g. $\Omega_0$ = $-$40 or $-$35 km s$^{-1}$ kpc$^{-1}$. However, we can preliminarily judge by eye that the spur signal revealed by our data is not present in most simulation, meaning that such a spur is not that easy to reproduce. Interestingly, some slower bars \citep{2024arXiv240207986H}, like (3.0, $-$25) or (5.1, $-$20), can generate extended structures at the same location as the spur in data, even though these streams present some irregular shapes that data do not show. Only when approaching the fiducial setting (5.5, $-$35) can models appear to be closer to the observation.

\begin{figure}[htb!]
\centering
\includegraphics[width=\columnwidth]{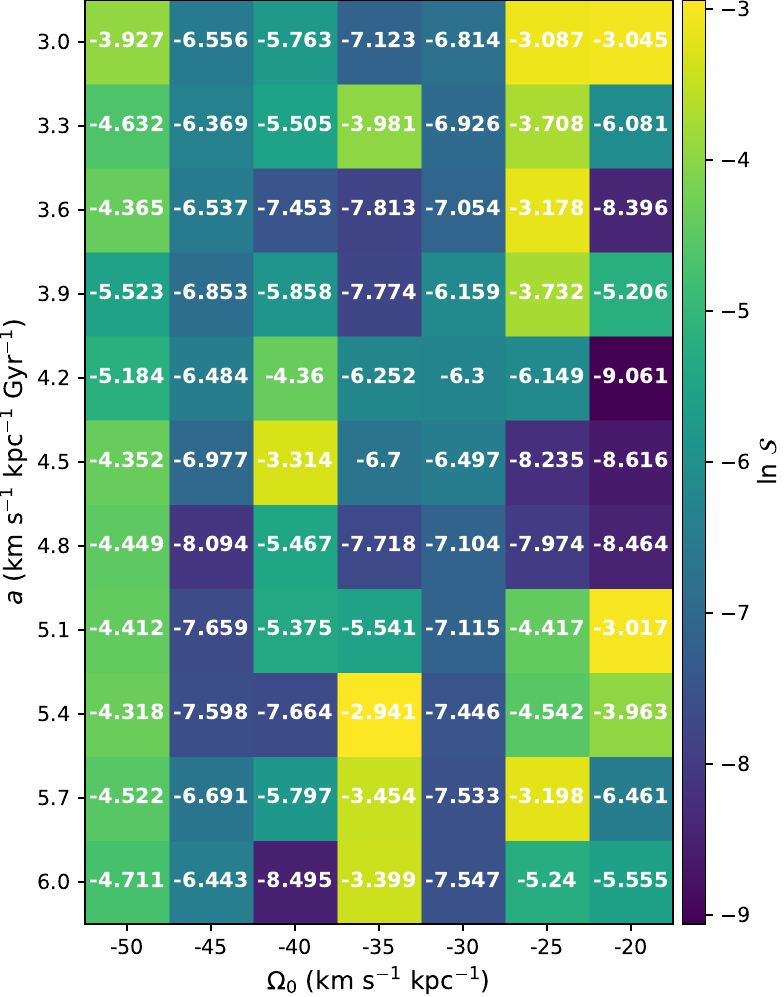}
\caption{ln$\mathcal{S}$ (Equation~\ref{eq:lns}) calculated at each grid point of ($a$, $\Omega_0$) in Figure~\ref{fig:explore_bar}.}
\label{fig:quantify_bar}
\end{figure}

Quantitatively, the ln$\mathcal{S}$ at each grid point of Figure~\ref{fig:explore_bar} is calculated and given in Figure~\ref{fig:quantify_bar}. The distribution of ln$\mathcal{S}$ is not unimodal because influences from the bar on the Ophiuchus are dramatic and a small variation of the bar might lead to a big change of the stream, being more consistent with the observation or probably less. Here the highest ln$\mathcal{S}$ = $-2.941$ goes to the grid point ($a$, $\Omega_0$) = (5.4, $-$35), the one closest to the fiducial setting (5.5, $-$35). Though fiducial ln$\mathcal{S}$ = $-3.064$ is a little lower, it still means that the consistency between our fiducial model and the data is at a relatively high level. There are other two grid points (3.0, $-$20) and (5.1, $-$20) having slightly higher ln$\mathcal{S}$ than the fiducial, but their fanning parts are too much spread compared to the data. Note that ln$\mathcal{S}$ value is grid-dependent, which means that the parameter pair with highest ln$\mathcal{S}$ will change if a denser or sparser grid is applied. Since we aim to see how different bars affect the stream, instead of using the stream to constrain the bar searching for the best model (which needs much more efforts and is beyond scope of this work), we end up with keeping ($a$, $\Omega_0$) = (5.5, $-$35) as the fiducial and exploring the specific grid above. 

\section{Discussion and Conclusion} \label{sec:summary}
In this work, we focused on the dynamical evolution of the Ophiuchus stellar stream. We use a Bayesian Mixture Model to identify the stream stars from $S^5$ data. Based on the data, we simulate the Ophiuchus by incorporating a slowing Galactic bar, and conduct a detailed analysis of the model stream.

Our scenario simultaneously explains the short length, spur and fanning features of the Ophiuchus stellar stream. This demonstrates the necessity of including a rotating bar when modeling a stream whose pericenter is rather close to the Galactic central region. Moreover, deceleration of the bar's pattern speed should be considered as well.

In this scenario, the Ophiuchus tidal tails are flipped (Figure~\ref{fig:snapshot}) because of torques from the bar (Figure~\ref{fig:history}), which accounts for the complicated morphology of the stream. Specifically, a proportion of its leading tail at the present day would have been on the trailing side if it were not torqued. In other words, we might see a ``fake'' leading tail today that originally had been the opposite trailing side in the past, and vice versa for a current trailing one (Figure~\ref{fig:slice}). Therefore, caution should be taken when investigating streams with small pericenters as their stars' orbits may have been redistributed by the bar. Concerning how to preliminarily judge whether this happens or not to a stream, we may observe distributions of its member stars in phase space. For example, firstly, the Ophiuchus stream has a complicated morphology including the spur and fanning, rather than being a simple linear structure. Secondly, in Figure~\ref{fig:model_vs_data}e, \ref{fig:validate}b or Figure~\ref{fig:bar_const_axi}b, the stream is not a single smooth sequence but rather is in a chaotic state in $\mu_\delta$, which might be a signature of its disordered evolutionary history. Besides these distributions, the stream is orbiting in a prograde manner (Figure~\ref{fig:pot}b), in the same as the Galactic bar, which makes it possible for them to interact for a sufficiently long time for the stream to be affected. This is similar to the case where streams should not only be close to the Large Magellanic Cloud but also have a small relative velocity (enabling longer interaction time) to experience the largest perturbation \cite{2021ApJ...923..149S}.

A gap or spur in a stream is usually attributed to perturbation from a flyby of a massive object \citep{2021ApJ...911..149L,2024arXiv240402953H}. However, it has been shown that the Galactic bar is able to reshape the density of the Palomar 5 stream \citep{2017MNRAS.470...60E} and even produce gap-like features \citep{2017NatAs...1..633P} (note that the Palomar 5 stream also has a prograde orbit). Here we also demonstrate the bar's ability to make a spur in the Ophiuchus stream. As a result, not just encounters with subhalos, but also the bar should be considered when looking for mechanisms explaining spurs in stellar streams, especially for streams orbiting prograde through the inner regions of the Milky Way.

The Ophiuchus is sensitive to the Galactic bar \citep{2016ApJ...824..104P,2016MNRAS.460..497H}. We also explore how the stream responds to a variety of the bar's rotations. By fixing the angle $\theta_0$ = $-30^\circ$, we show that the observation of Ophiuchus qualitatively (Figure~\ref{fig:explore_bar}) and quantitatively (Figure~\ref{fig:quantify_bar}) favors the model stream obtained when the bar approximates the fiducial setting where the current pattern speed $\Omega_0$ $=$ $-35$ km s$^{-1}$ kpc$^{-1}$ and the deceleration $a$ $=$ 5.5 km s$^{-1}$ kpc$^{-1}$ Gyr$^{-1}$ by assuming a uniform deceleration (Equation~\ref{eq:uniform}), which closely follows the evolution of the pattern speed from \citet{2021MNRAS.500.4710C} over the past 3 Gyr (Figure~\ref{fig:pot}c). 

In a wider context, our results have offered a different viewpoint for understanding the deceleration of the Galactic bar from the perspective of the Ophiuchus stream. If our Galaxy is typical, it is even possible to provide crucial constraints to the evolution of bars in galaxy formation models by studying the Milky Way bar through the Ophiuchus stellar stream.

Finally, we have made three animations to visualize the evolution of the young, mid, and old groups of the Ophiuchus stream in our barred model, along with another animation as a comparison illustrating the stream in an axisymmetric potential, which can be retrieved at \url{https://doi.org/10.5281/zenodo.15003380}.

\section{Acknowledgments}

Y.Y., G.F.L., S.L.M., and D.B.Z. acknowledge support from the Australian Research Council through Discovery Program grant DP220102254. 
This work is part of the ongoing $S^5$ (\url{https://s5collab.github.io}). 
T.S.L. acknowledges financial support from the Natural Sciences and Engineering Research Council of Canada (NSERC) through grant RGPIN-2022-04794.
S.A.U. acknowledges support from the American Association of University Women (AAUW) through their American Dissertation Fellowship. S.E.K. acknowledges support from the Science \& Technology Facilities Council (STFC) grant ST/Y001001/1. This paper includes data obtained with the Anglo-Australian Telescope in Australia. We acknowledge the traditional owners of the land on which the AAT stands, the Gamilaraay people, and pay our respects to elders past and present.

This work presents results from the European Space Agency (ESA) space
mission Gaia. Gaia data are being processed by the Gaia Data
Processing and Analysis Consortium (DPAC). Funding for the DPAC is
provided by national institutions, in particular the institutions
participating in the Gaia MultiLateral Agreement (MLA). The Gaia
mission website is \url{https://www.cosmos.esa.int/gaia}. The Gaia archive
website is \url{https://archives.esac.esa.int/gaia}.

This project used public archival data from the Dark Energy Survey
(DES). Funding for the DES Projects has been provided by the
U.S. Department of Energy, the U.S. National Science Foundation, the
Ministry of Science and Education of Spain, the Science and Technology
Facilities Council of the United Kingdom, the Higher Education Funding
Council for England, the National Center for Supercomputing
Applications at the University of Illinois at Urbana-Champaign, the
Kavli Institute of Cosmological Physics at the University of Chicago,
the Center for Cosmology and Astro-Particle Physics at the Ohio State
University, the Mitchell Institute for Fundamental Physics and
Astronomy at Texas A\&M University, Financiadora de Estudos e
Projetos, Funda{\c c}{\~a}o Carlos Chagas Filho de Amparo {\`a}
Pesquisa do Estado do Rio de Janeiro, Conselho Nacional de
Desenvolvimento Cient{\'i}fico e Tecnol{\'o}gico and the
Minist{\'e}rio da Ci{\^e}ncia, Tecnologia e Inova{\c c}{\~a}o, the
Deutsche Forschungsgemeinschaft, and the Collaborating Institutions in
the Dark Energy Survey.  The Collaborating Institutions are Argonne
National Laboratory, the University of California at Santa Cruz, the
University of Cambridge, Centro de Investigaciones Energ{\'e}ticas,
Medioambientales y Tecnol{\'o}gicas-Madrid, the University of Chicago,
University College London, the DES-Brazil Consortium, the University
of Edinburgh, the Eidgen{\"o}ssische Technische Hochschule (ETH)
Z{\"u}rich, Fermi National Accelerator Laboratory, the University of
Illinois at Urbana-Champaign, the Institut de Ci{\`e}ncies de l'Espai
(IEEC/CSIC), the Institut de F{\'i}sica d'Altes Energies, Lawrence
Berkeley National Laboratory, the Ludwig-Maximilians Universit{\"a}t
M{\"u}nchen and the associated Excellence Cluster Universe, the
University of Michigan, the National Optical Astronomy Observatory,
the University of Nottingham, The Ohio State University, the OzDES
Membership Consortium, the University of Pennsylvania, the University
of Portsmouth, SLAC National Accelerator Laboratory, Stanford
University, the University of Sussex, and Texas A\&M University.
Based in part on observations at Cerro Tololo Inter-American
Observatory, National Optical Astronomy Observatory, which is operated
by the Association of Universities for Research in Astronomy (AURA)
under a cooperative agreement with the National Science Foundation.


%

\vspace{5mm}
\facilities{Anglo-Australian Telescope (AAOmega+2dF)}


\software{rvspecfit \citep{2019ascl.soft07013K},
astropy \citep{2013A&A...558A..33A},
gala \citep{2017JOSS....2..388P},
emcee \citep{2013PASP..125..306F},
AGAMA \citep{2019MNRAS.482.1525V},
dustmaps \citep{2018JOSS....3..695G},
numpy \citep{harris2020array},
matplotlib \citep{Hunter:2007}
}





\bibliography{sample631}{}
\bibliographystyle{aasjournal}



\end{document}